# Exploring Molecular Odor Taxonomies for Structure-based Odor Predictions using Machine Learning

Akshay Sajan,[a] Stijn Sluis,[a] Reza Haydarlou,[a] Sanne Abeln,[b] Pasquale Lisena,[c] Raphael Troncy,[c] Caro Verbeek,[d] Inger Leemans,[e] and Halima Mouhib[a]

[a] Department of Computer Science, VU Bioinformatics Group, Vrije Universiteit Amsterdam, De Boelelaan 1105, 1081 HV Amsterdam, The Netherlands.
[b] Department of Computer Science, AI Technology for Life, Universiteit Utrecht, Heidelberglaan 8, 3584 CS Utrecht, The Netherlands
[c] EURECOM, Campus SophiaTech, 450 Route des Chappes, 06410 Biot, France.
[d] Faculty of Humanities, Art and Culture, History, Antiquity, De Boelelaan 1105, 1081 HV Amsterdam, The Netherlands
[e] KNAW Humanities Cluster, Oudezijds Achterburgwal 185, 1012 DK Amsterdam, The Netherlands. Vrije Universiteit Amsterdam, Department of Art and culture, History, and Antiquity, Faculty of Social Sciences and Humanities, De Boelelaan 1105, 1081 HV Amsterdam, The Netherlands.

**Abstract**

One of the key challenges to predict odor from molecular structure is unarguably our limited understanding of the odor space and the complexity of the underlying structure-odor relationships. Here, we show that the predictive performance of machine learning models for structure-based odor predictions can be improved using both, an expert and a data-driven odor taxonomy. The expert taxonomy is based on semantic and perceptual similarities, while the data-driven taxonomy is based on clustering co-occurrence patterns of odor descriptors directly from the prepared dataset. Both taxonomies improve the predictions of different machine learning models and outperform random groupings of descriptors that do not reflect existing relations between odor descriptors. We assess the quality of both taxonomies through their predictive performance across different odor classes and perform an in-depth error analysis highlighting the complexity of odor-structure relationships and identifying potential inconsistencies within the taxonomies by showcasing pear odorants used in perfumery. The data-driven taxonomy allows us to critically evaluate our expert taxonomy and better understand the molecular odor space. Both taxonomies as well as a full dataset are made available to the community, providing a stepping stone for a future community-driven exploration of the molecular basis of smell. In addition, we provide a detailed multi-layer expert taxonomy including a total of 777 different descriptors from the Pyrfume repository.

# 1 Introduction

Despite numerous rules of thumb proposed by fragrance chemists over the years, predicting the smell of molecules directly from molecular structure remains a challenging task[1]. One of the main goals in fragrance chemistry is to relate the physical and structural properties of odorants to their perceptual odor properties in order to rationally design new compounds of a desired smell. The complexity of predicting smell directly from structure arises from the fact that molecules with similar structures can possess very distinct odors, while structurally unrelated molecules with different structures may produce close to identical odors[2]. Figure 1 shows selected examples of odorant molecules (odorants) and their reported odor descriptors. In the different cases, it can be seen that the addition of a simple methyl group (depicted in blue) has different effects on the smell of the original odorant, ranging from no effect to a complete loss of the biological activity.

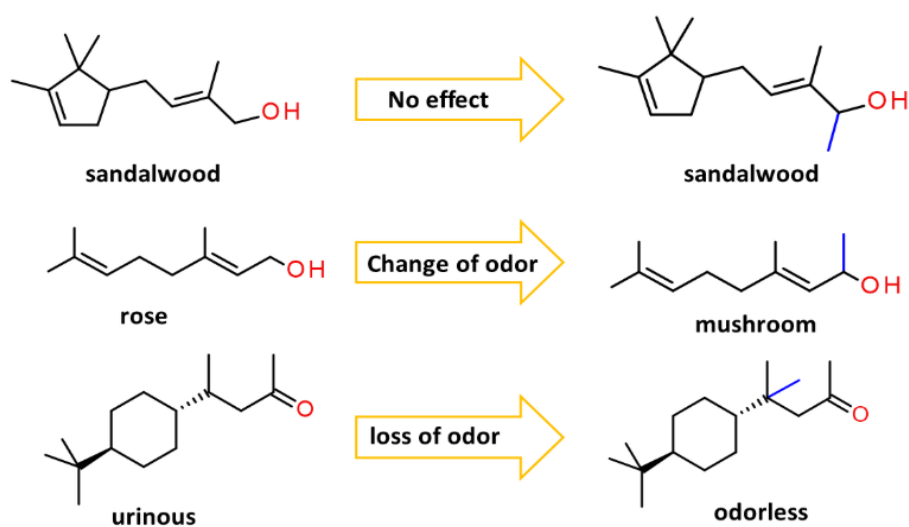

**Figure 1**. Impact of a small structural modification (addition of a single methyl group, symbolized by blue line). The effect ranges from no change to a complete loss of odor. Figure adapted from ref.[2].

In addition to this and the complex underlying biology, describing a smell and quantifying it directly depends on semantic language and individual perception of trained experts. This is a different situation from other senses such as vision and hearing that do not necessitate subjective input. Nonetheless, the overall complexity of the sense of smell provides a unique opportunity to explore the usefulness of machine learning techniques, for instance to identify relevant structural patterns directly from available molecular data without the necessity to elucidate the complex biology. While several machine learning approaches have been reported for this purpose over the past decade[3–5], the most recent breakthrough was achieved using a graph neural network (GNN) to generate an embedding for molecular representation which is plotted as Principal Odor Map (POM)[6]. Using this GNN-derived embedding space allowed for accurate structure-based odor predictions of novel odorants that outperformed median human panelists. Despite this success, it is still difficult to discern which facets of the molecular structure contribute to the odor classification task due to the limited interpretability of deep learning technologies. Still, the POM provides additional insight into the underlying perceptual odor similarity between odor descriptors and suggests an intrinsic conceptual hierarchy between the descriptors that may be explored to build a taxonomy. Therefore, we hypothesize that structuring odor descriptors into a taxonomy that captures hierarchical and correlated relationships can enable machine learning models to learn more robust and generalizable patterns from molecular-odor data. This concept is visualized in Figure 2, where our proposed odor taxonomy is applied in the classification of two popular molecular examples from perfumery and the flavor industry: geraniol and grapefruit mercaptan.

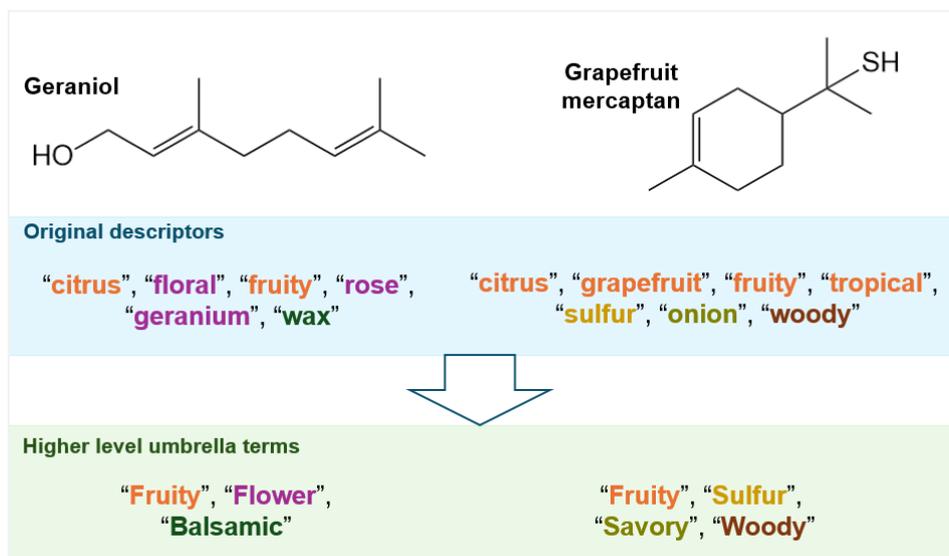

**Figure 2.** Example of applying the expert taxonomy on geraniol and grapefruit mercaptan. By introducing the taxonomy, the output space is reduced from six to three and seven to four descriptors for geraniol and grapefruit mercaptan, respectively. The taxonomy is imposed upon the dataset by replacing the odor descriptor annotations of each molecule with their respective higher level umbrella terms (descriptors falling under a higher level umbrella term are color coded with matching colors). Note that in the case of grapefruit mercaptan the smell impression gradually shifts from fruity to sulfur within increasing concentration. This dosage-dependent effect is not taken into account for the ML classifiers trained in this work.

Although simple taxonomies based on clustering odor descriptors on co-occurrence from structure-odor data via web scraping have been reported before,[7] there is no available open source taxonomy for molecular descriptors that can be used to improve structure–odor prediction by organizing odor descriptors into semantically or perceptually meaningful groups. To address this gap, we use a selection of several datasets available on the Pyrfume repository[8] that comprises more than 40 stimulus-linked olfactory datasets across different species. In total, Pyrfume provides a selection of over 20,000 odorants and over 770 distinct odor descriptors. However, the repository also contains data that was obtained through web scraping and needs to be carefully checked before usage. For this work, we selected all the useful datasets from the Pyrfume collection and aggregated a molecular dataset that can be used to explore structure-based odor prediction using machine learning models. Thus, we investigate whether the hierarchical ordering of odor descriptors helps convey essential information that allows us to deepen our understanding of structure-odor relationships. Our work provides three main deliverables, namely (i) a compiled molecular dataset for structure-based odor prediction tasks, (ii) an expert and a data driven taxonomy for odor descriptors; and (iii) insight into model prediction and interpretability by way of linking molecular structural patterns to odor profiles. All the data, taxonomies, and models are openly accessible on the GitHub repository (see section 6). We provide an improved predictive performance of interpretable machine learning models for structure-based odor predictions based on an expert and a data-driven odor taxonomy. Both taxonomies improve the predictions of Logistic Regression, random forest, and XGBoost[9] models, and outperform random groupings of odor descriptors that do not reflect existing relations between odor descriptors. We evaluated the quality of both taxonomies based on their predictive performance across diverse odor classes and conducted a detailed error analysis that reveals the underlying complexity of odor–structure relationships. Finally, we discussed current limitations, challenges, and future opportunities and directions of structure-based odor predictions using machine learning. Although there are more aspects of the olfaction sense where data has been gathered to leverage machine learning techniques, for instance in the physiology of odor recognition or genetic patterns in olfactory phenotypes, our work focuses on the task of structure-based odor prediction. For other purposes, the interested reader may consult several reviews that have treated these topics[10,11].

## 2 Methods

In order to circumvent the complex biology and leverage conceptual hierarchies between odor descriptors to improve the performance of interpretable machine learning models, we build two different taxonomies: an expert-derived taxonomy and a data-driven taxonomy based on the co-occurence of odor descriptors within the molecular dataset. In our approach, we built a curated structure-odor dataset, containing expert-annotated odor descriptors. The odor descriptors from the merged datasets are used as a basis to build the taxonomies that are subsequently used to benchmark the predictive performance of interpretable machine models. A birds eye view of the approach is depicted in Figure 3.

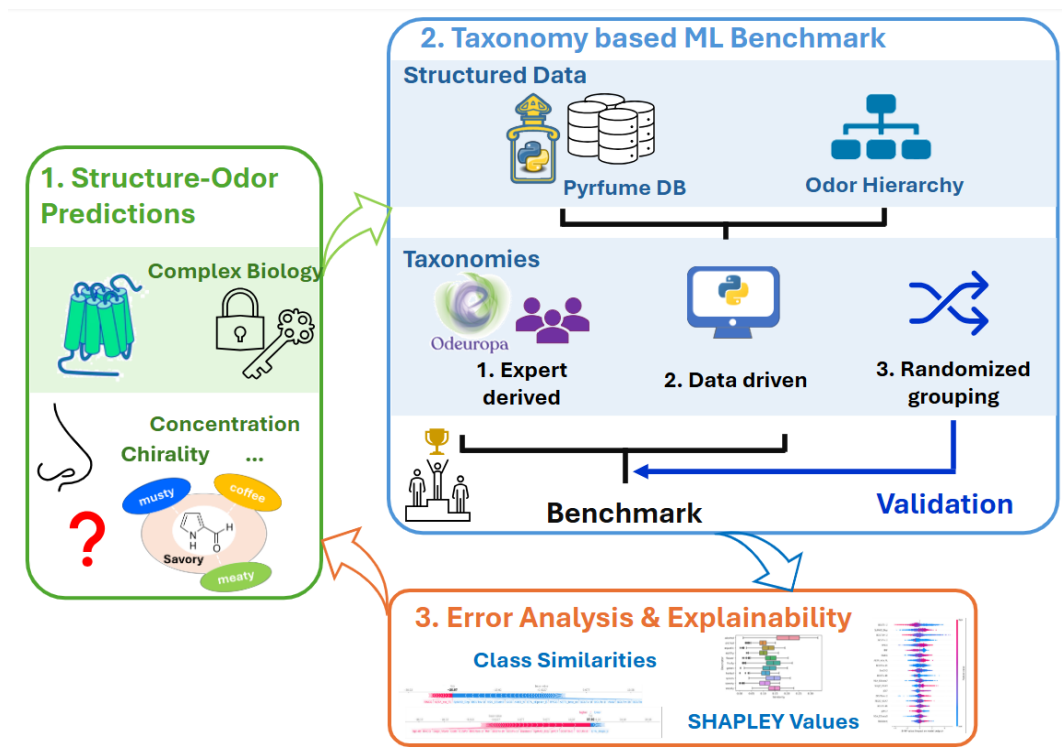

**Figure 3.** Schematic overview of the implementation used in this work. Expert and data driven taxonomies are leveraged to improve machine learning based odor prediction from openly accessible molecular datasets. We use randomized grouping of the 146 unique descriptors to benchmark the usefulness of the two taxonomies. Subsequently, the predictive performance of both taxonomies across different odor classes serves as a quality metric that allows for an in-depth error analysis which highlights the inherent complexity of odor-structure relationships.

### 2.1 Dataset preparation

The molecular datasets used in this study were assembled using the open access repository Pyrfume. Pyrfume is an open-source project aimed at analyzing odorants and their features, providing both tools and access to diverse datasets. These datasets include molecular structures, perceptual data, and physicochemical properties, making it a valuable resource for researchers in the field.

From around fifty datasets in the repository, 14 datasets contained data about odor descriptors and compounds : Arctander[12], aromaDB[13], Dravnieks, FooDB[14], FlavorDB[15], Flavornet[16], Goodscents[17], IFRA[18], Keller (2012[19] and 2016[20]), National Geographic's Smell Survey[21],

Sharma (2021a[22] and 2021b[23]), Sigma Fragrance and Flavor Catalog (2014)[24], Snitz[25] and Leffingwell[26].

Among these, 7 datasets were selected to be used for the classification task: Arctander, aromaDB, FlavorDB, Flavornet, Goodscents, IFRA, and Leffingwell. The selection was based on: 1) datasets that concern human subjects (excluding animal based datasets), 2) datasets that contain descriptors for single odorants (molecules, not mixtures). Furthermore, these are the most relevant datasets with respect to the reliability of the reported odorants, their odor descriptors, and listed physicochemical properties. The datasets were merged and processed to create a comprehensive dataset with 6711 molecules and 146 distinct odor descriptors. In the remainder of this paper, we will refer to this aggregation as **Merged molecular dataset (MMD)**. Table 1 provides a summary of the merged datasets, the publication year, the number of chemical compounds, and the number of odor descriptors.

**Table 1.** Summary of the datasets retrieved from the Pyrfume repository to (i) build the molecular dataset (comprising a total of 6711 different molecules with 146 unique odor descriptors) used to train different ML classifiers, and (ii) merge a maximum number of odor descriptors for the open access expert taxonomy, which now comprises a total of 557 unique odor descriptors and 60 hedonic odor qualities (see section 2.2.1 for additional details).

| dataset | Publication year | # Compounds | # Odor Descriptors |
|---|---|---|---|
| **Arctander**[12] | 1960 | 2580 | 762 |
| **AromaDB**[13] | 2018 | 869 | 127 |
| **FlavorDB**[15] | 2018 | 525 | 255 |
| **Flavornet**[16] | 2004 | 716 | 195 |
| **Goodscents**[17] | 2004 | 4622 | 667 |
| **IFRA**[18] | 2019 | 1146 | 191 |
| **Leffingwell**[26] | 2001 | 3487 | 113 |
| **MMD**[1,2)] | This work | 6711 | 146 |
| **Expert Taxonomy (ET)** | This work | n.a. | 617 |

[1)] The number of compounds is not an exact addition over all datasets due to the overlap of samples that needed to be removed to avoid duplicates. Henceforward, this aggregation is referred to as **Merged molecular dataset (MMD)**
[2)] Note that the **data-driven taxonomy (DT)** is based on the 146 odor descriptors of the MMD

Selected results from the initial data exploration, e.g., distribution of molecular weights, number of descriptors per molecular sample, and general statistics of the dataset are provided in Figures S4-S11. The majority of the compounds possess 1 to 5 labels, while 32 compounds exhibit 13 odor labels or more descriptors, with a maximum of 17 descriptors. It should be noted that for each molecule the strength of the individual odor descriptors is not provided and the descriptors are considered to be equally strong. In addition, the data is imbalanced towards "nice" smells that are linked to the food and perfume industry. There are more instances within the fruity class and floral class of the training dataset, 1982 and 1410 respectively, then in the camphor class with only 215 molecular samples.

To train and validate the machine learning models (see section 2.3 for details), the dataset was split into a training and test dataset using a second-order iterative stratification to take the higher-order relationships between labels into account when doing a data split and make sure that the distribution of label pairs between splits stays consistent. The class distribution of the datasets is consistent across splits and is the case as well after imposing both taxonomies as can be seen in Figures S9-S10.

## 2.2 Taxonomy derivation

To leverage the hierarchical connections within the odor descriptors for the machine learning classification, we rely on two different approaches: benefiting from expert domain knowledge by manually going through all 617 odor descriptors across the 14 previously-mentioned datasets in Pyrfume and grouping them into classes that reflect existing relations between them, as well as generating a data-driven correlation-based clustering directly from the MMD. The detailed process to compile both taxonomies is described below.

### 2.2.1 Expert-derived multi-domain taxonomy of odor descriptors (ET)

In order to have a complete list of all descriptors, we aggregated descriptors coming from all 14 datasets if they had an identical label or if the Levenshtein distance between labels was <1. This enabled us to consider as a single concept small differences in spelling like 'fish' and 'fishy' or 'wine-like' and 'winelike'. The aggregations have been manually validated by experts to avoid any possible mismatch. This aggregation leads to 617 distinct descriptors. Most of these descriptors are present in only one source dataset.

Our team possesses expertise in chemistry, perfumery and historical smell culture, which allows us to propose a taxonomy to group the gathered 617 odor descriptors in different classes based on our domain knowledge. We decided to indicate descriptor groups: (i) source-based descriptors: descriptors relating the smell to a single odor (lemonlike, yeasty, wormwood) or overarching class of scents (fruity, woody, green) [557 concepts] and (ii) olfactory qualities: general adjectives expressing hedonic, trigeminal or emotional responses, or intensifiers (nice, fresh, fragrant, putrid, dry, light, heavy) [60 concepts].

The Source-based descriptors were divided in 16 different classes 'scent families' and 31 subclasses, e.g., the "Alcohol" class contains the sub-classes "Acid" and "Alcohol", which each contains their own odor descriptors. The list of classes and sub-classes are provided in the supporting information, the hierarchical expert taxonomy is provided on the [ODEUROPA website](#).

Henceforward, the taxonomy will be referred to as Expert derived Taxonomy (ET). For this paper we only consider the 557 source-based descriptors, since the quality descriptors are harder to relate to specific odorant molecules. General statistics of the dataset after imposing the expert dataset are provided as supporting information.

### 2.2.2 Data driven taxonomy (DT)

Next to the expert taxonomy, it is useful to generate a data-driven taxonomy directly from co-occurence of descriptors within the MMD. This allows to avoid (or identify) bias from experts and can help to later finetune the expert taxonomy.

The odor descriptors of the MMD co-occur frequently with perceptually similar odor descriptors. A correlation matrix based on this co-occurrence was used to group the odor descriptors into 16 clusters using Agglomerative hierarchical clustering, a clustering method that iteratively merges individual descriptors to form groups with minimal internal distances, Euclidean in our case, as well as using Ward linkage to minimize inter-cluster variance (see ESI for parameters used). The number of clusters was set to 16 to allow the comparison with the ET, where 16 classes are used to group conceptually related odor descriptors. It should be noted, that while the data-driven taxonomy has been intentionally limited to 16 classes, the elbow method suggests an ideal number of 43 groupings based on co-occurence, which may be used to generate a co-occurrence based hierarchical multi-domain taxonomy in future work. Figure 4 depicts a zoom into 19 odor descriptors of the correlation matrix, based on the co-occurrence of the odor descriptors within the training split of MMD. The figure shows two main clusters that can intuitively be attributed to an "alcohol" and a "fruity" odor cluster. Note that the odor descriptors "wine" and "fruity", as well as "apricot" and "peach" have a high frequency of co-occurence, which would also be an intuitive association to most people. The full correlation matrix containing all 146 descriptors is provided as supporting information (Figure S4). Henceforward this taxonomy will be referred to as Data driven Taxonomy (DT). The complete list of all clusters is provided in the supplementary materials. In the discussion, the 16 classes from both taxonomies are compared against one another to elucidate the choices made by the experts compared to the clusters revealed directly from the co-occurrence matrix of odor descriptors.

### 2.3 Machine learning model, feature selection and hyper-parameter tuning

To evaluate whether the grouping of higher order odor descriptors helps the machine learning task, we compare the performance of machine learning models applied on the MMD using both taxonomies (ET and DT). Three interpretable machine learning models were selected to benchmark the usefulness of taxonomies for structure-based odor predictions: Logistic regression, Random Forest, and XGBoost[9]. The hyperparameter tuning of the model was done using Bayesian optimization through skopt[27]. The final features and hyperparameters used are provided in the supporting information (Table S1-S3). Modred, a free open-source molecular descriptor calculator[28], was used to generate molecular features from the simplified molecular input line entry system (SMILES) of all the molecules in our datasets. The resulting descriptors consist of a total of 713 molecular features, after removing columns with missing values, ranging from atom types, number of specific atom types, acid/base counts, as well as older obscure chemoinformatic descriptors like Burden matrices and Chi-descriptors. The derived molecular descriptors can consist of features using topological representations (2D) and geometrical representations (3D). The generated features then underwent an exhaustive feature selection procedure. 0 variance features were removed and then ANOVA F values were used to filter the features and the best feature for each of the 146 classes was selected.

This was then narrowed down further through recursive feature elimination, using a random forest with Permutation Feature Importance, the features are narrowed down to 23.

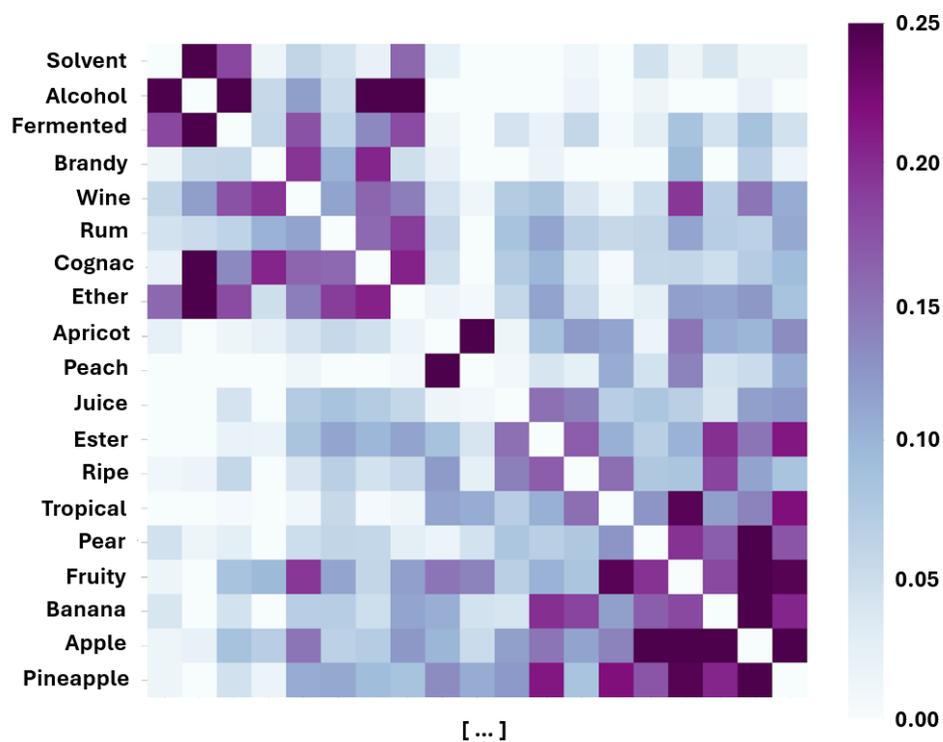

**Figure 4.** Selection of the correlation matrix based on the co-occurrence of 19 odor descriptors of the MDD. The full correlation matrix is provided as supporting information (Figure S4). The indices are sorted based on the clustering made using hierarchical agglomerative clustering with 'ward' linkage and using Euclidean distances. A cluster of labels that often co-occur appears as a dark area on the heatmap. The data-driven taxonomy leverages these co-occurrence patterns to generate the classes and group the descriptors based on the occurrence throughout the data (see methods section for additional details).

## 2.4 Randomized grouping of odor descriptors

In order to validate our approach and benchmark the predictive performance of the proposed taxonomies, we generated 1000 randomized groupings of all odor descriptors to provide taxonomies that group odors randomly, independently from any conceptual similarities. This validation step is crucial to guarantee that performance increase is not due to reducing the complexity of the classification task, but also because the model catches upon existing links that can help to interpret structure-odor correlation. Therefore, the odor descriptors are randomized throughout the 16 classes of the ET and DT. Hereby, the randomization maintains the template of the original taxonomical structure, i.e., the number of odor descriptors in a given class (see Table S8-S9).

For each randomized grouping, the performance of the machine learning models are stored and their scores are assessed through the macro AUROC, F1, Precision and Recall. Due to computational constraints, the features and hyperparameters derived from the original descriptor dataset, in the previous section, are used here.

Permuting all 146 descriptors within their respective umbrella terms yields 146 factorial (146!) possible combinations. While this is not computationally feasible, an approximation of the score distribution, from the reduction of the output space for all metrics, can be obtained using the central limit theorem. The central limit theorem states that with the increase in sample size, the sampling distribution converges toward the population distribution[29]. The randomization sampling has been limited to 1000 trials for both taxonomies. To ascertain if the distribution of all metrics have converged, we plot the running mean and standard distribution for all metrics. This is checked for the score metrics for both taxonomies as well as their summations since there might be differences due to variations in taxonomical structure. We use this combined distribution for all metrics as representative of the increase in performance for all metrics due to the reduction of the output space. We then validate both taxonomies by comparing the metric scores against this distribution: if the taxonomies outperform the distribution, it would imply that there is more to the performance gain than just the reduction of the output space.

## 2.5 Interpretability

To investigate the most relevant features that the machine learning model uses for its predictions we investigated the feature importance of the different models. Firstly, the permutation feature importance (PFI) was used to assess the impact of each feature on the model performance. Here, permutations are added randomly to the features of the test data to rank features by measuring the decrease of a given score metric. A key limitation of PFI is the lack of class-specific insight it can provide as it focuses mainly on the overall macro scores. To circumvent this, we subsequently carried out a SHapley Additive exPlanation (SHAP) value analysis[30]. This method makes use of cooperative game-theory to assign a value to each feature representing its contribution to the classification decision of the model. The SHAP values are plotted in summary plots for each class to understand the feature contributions for each class for both taxonomies to interpret the model output and the decision making that went behind it.

## 3 Results and Discussion

### 3.1    Taxonomy validation through randomized grouping of odor descriptors

To evaluate the predictions using the two structured taxonomies, we first established a baseline model using the full unstructured set of odor descriptors. This initial step allows us to assess the inherent difficulty of predicting individual odors directly from molecular features, and to quantify the benefit of using structured hierarchical taxonomies. The baseline model performance of the XGBoost classifier trained to perform multi-class classification over 146 classes (using the MMD) achieved an overall score of 0.6 AUC, which is above random and suggests that the model is able to capture some meaningful signal in the data. However, this performance remains modest indicating that this is not a trivial classification task. There are several potential limitations that can lead to this such as difficult feature representation, the limited data quality, and potentially difficult class separability based on underlying structural patterns in the data. Despite extensive hyperparameter tuning, none of the models yielded performances as reported in previous work[5]. We thus use this model as a baseline and compare it to the performance for the different models when the derived taxonomies are imposed and classification task is simplified to 16 classes.

The comparative evaluation of the different performances using an XGBoost classifier are provided in Table 3 and Figure 4. The results show that models trained on the dataset with imposed taxonomies outperform models trained on the original descriptors across all metrics for all the classifiers. Both taxonomies perform better than the distribution of the performance metrics of the randomized groupings of odor descriptors. It should be noted that the performance metrics of the randomized taxonomies is higher than 0.5 which would mean that the models are still picking up on a signal or structure in the data despite the loss of meaningful class groupings from the taxonomies. This is because the randomized taxonomies, for the most part, are not entirely meaningless due to the intercorrelation of many odor descriptors. This demonstrates the inherent complex hierarchical relationships of the odor descriptor space and the limitations of a two-layer taxonomy to capture conceptual links between different descriptors.

Despite the high number of possible combinations for the randomized groupings (146! (factorial) in total), it is almost unavoidable to create classes where none of the descriptors are correlated. The scores for the randomized groupings can serve as a baseline reflecting the ability of the model to extract signal from noise and serve as a proxy for the precision of ET and DT. Hereby, a higher score indicates a better conceptual clustering and arrangement of the odor descriptors. Both taxonomies outperform the random groupings, with the DT showing a marginal advantage over the expert taxonomy.

It should be noted that the reported performances represent the average over all 16 classes. For the following sections, we focus on the performance results obtained using the XGBoost classifier, which outperformed the logistic regression and the random forest classifiers. The performances of the two latter models, as well as the class-wise scores of the XGBoost model for both taxonomies are provided in the ESI (Tables S4-S7).

**Table 2.** Performance metrics of the XGBoost classifier as obtained using the full 146 descriptors, the taxonomies, and the randomized grouping. The results are visualized in Figure 4. The corresponding performance metrics for the random forest and logistic regression models are provided as supporting information.

| Dataset | AUROC | F1 | Precision | Recall |
|---|---|---|---|---|
| **MMD** | 0.604 | 0.268 | 0.393 | 0.219 |
| **ET**[1] | 0.684 | 0.496 | 0.553 | 0.455 |
| **DT**[1] | 0.698 | 0.513 | 0.567 | 0.474 |
| **Random grouping**[2] | 0.648(07) | 0.455(15) | 0.511(15) | 0.418(15) |

[1] The classwise performance metrics are provided in the ESI (TableS6 and TableS7).
[2] The standard deviation for the 1000 randomized grouping is given in parentheses with respect to the last 2 digits (see ESI for additional details).

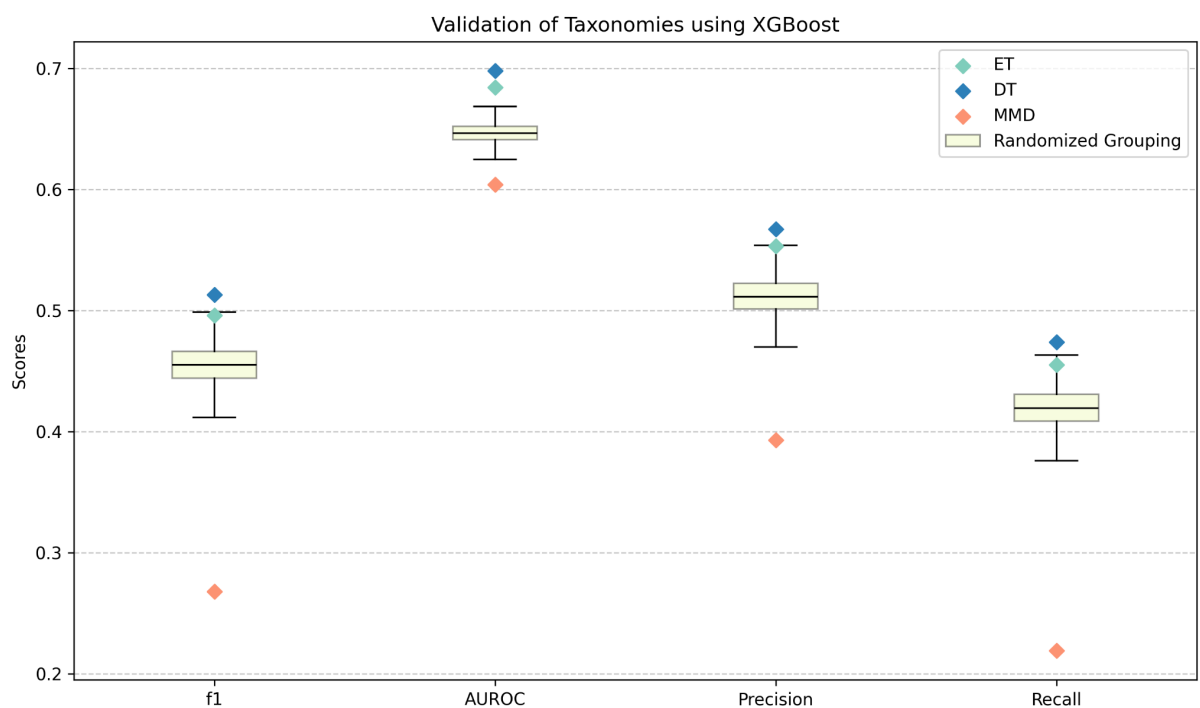

**Figure 5**. Box plot comparing the performance metrics of the XGBoost classifier using the randomized grouping and the two benchmarked taxonomies (ET and DT) as well as the scores obtained from the original descriptors. The numerical values of each metric are provided in Table 2.

**3.2 Conceptual comparison of derived taxonomies**

To provide insight into how well each taxonomy reflects meaningful relationships between odor descriptors and how these align with known concepts in perfumery and sensory science, it is essential to go beyond the quantitative metrics-based comparison and include semantic coherence, domain knowledge, and structural elements of the taxonomies. Therefore, we compare the 16 data-driven groups (Classes A to P) to the 16 expert-derived classes to identify conceptual similarities between them. The complete list of odor descriptors and the different odors grouped together to build the taxonomies are provided in the supporting information (see Tables S8 and S9). While this comparison highlights where the two taxonomies overlap and differ from one another, it also raises the question how easily each grouping of odor descriptors may be interpreted by a human reader or by a large-language model (LLM). To address this, we used the latest version of ChatGPT (GPT-4o) to define class names for the 16 classes within the DT (The prompt used to generate the classes is provided in the ESI). The comparison is shown in Table 3. While several classes of the DT can be directly mapped to the classes defined within the expert taxonomy, i.e., "Gourmand", "Flower", "Alcohol", "Savory", "Fruity", "Sulfur", and "Woody", the others are more difficult to assess. The classes that can be directly mapped to expert defined classes reflect groupings of odor descriptors that are mainly based on conceptual odor similarities, which was the central element to structure the expert taxonomy. Other data-driven classes, e.g., class "E" cannot be directly mapped to one of the 16 expert defined classes. Class "E" contains a mix of conceptually related odor descriptors such as 'cinnamon', 'clove', 'spicy', and 'vanilla' which may be grouped under the higher term 'spicy'. However, it also includes seemingly unrelated terms such as 'balsamic', 'medicinal', 'phenol', 'smoked'. These descriptors are conceptually not similar to the listed 'spicy' descriptors, but rather evoke medicinal practices or properties. Vanilla is also part of the gourmand family within perfumery and sometimes classified as having a green aspect[31]. The DT derived class "E" can therefore rather be described as a "Health" or "Medicinal" category. More in alignment with the ET that shows inherent bias towards grouping odor descriptors on conceptual similarities, the suggestion from the ChatGPT describes class "E" as "spicy" without taking the outliers 'balsamic', 'medicinal', 'phenol', 'smoked' into account. It should be noted that "medicinal" and "smoky" do not seem source based, whereas the other descriptors do. Often myrrh, which is a resin, is qualified as medicinal[32].

**Table 3.** Comparison of the 16 odor classes within the 2 defined taxonomies, including the number of descriptors within each class. The class definitions for the DT groupings (Classes A to P) are compared to the ET categories and to categories provided through a large language model (LLM: GPT-4). Class descriptions that are similar between the two taxonomies are highlighted in bold. The full list of descriptors within the different classes of the ET and DT are provided as supporting information (Table S8 and S9).

| Class[1] | # descriptors[2] | Class[3] | # descriptors[4] | Similarity[5] | LLM[6] |
|---|---|---|---|---|---|
| **Alcohol** | 10 | A | 12 | Anethole | Aromatic |
| **Animal** | 4 | B | 14 | **Gourmand** | **Gourmand** |
| **Aquatic** | 3 | C | 22 | **Flower**[7] | **Floral** |
| **Balsamic** | 2 | D | 13 | **Pungent** | **Pungent** |
| **Chemical** | 12 | E | 8 | [8] Health/Medicinal | Spicy |
| **Earthy** | 5 | F | 8 | [8] **Dairy/Cream** | **Lactonic** |
| **Flower** | 15 | G | 8 | [8] **Green**/Fat | **Green** |
| **Fruity** | 28 | H | 8 | **Alcohol** | **Alcoholic** |
| **Gourmand** | 13 | I | 7 | **Savory** | Animalic |
| **Green** | 9 | J | 11 | **Fruity** | **Fruity** |
| **Herbal** | 5 | K | 8 | [8] Fresh | Citrus |
| **Savory** | 23 | L | 6 | [8] Clean | Camphorous |
| **Smoky** | 5 | M | 4 | **Acid** | **Acidic** |
| **Spices** | 8 | N | 3 | **Sulfur** | **Sulfurous** |
| **Sulfur** | 3 | O | 10 | **Woody** | **Woody** |
| **Woody** | 9 | P | 4 | **Berry** | **Berry** |

[1] 16 classes defined by domain experts for the expert-derived taxonomy.
[2] Number of descriptors included in each class of the expert-derived taxonomy.
[3] 16 classes obtained from the data driven taxonomy.
[3] Number of descriptors included in each class of the data driven taxonomy.
[5] Conceptual similarity of the data driven classes to the expert defined classes as estimated from the experts' perspectives.
[6] Suggested class name for the descriptors in the different classes of the data taxonomy as suggested using the large language model (LLM) ChatGPT-4o. See supporting information for the list of descriptors within the different classes.
[7] In perfumery this class is usually called 'floral'.
[8] Note that not all the classes from data taxonomy can be mapped directly to classes within the expert-derived taxonomy. The highlighted classes are not fully based on conceptual similarities but also evoke effect or odor practices (see text for more details).

There are four other DT derived classes that exhibit similar patterns: classes 'F', 'G', 'K', and 'L'. Looking closer into the descriptors of class 'K' and class 'L', the overarching name for these classes may be set to "Fresh" and "Clean", while the LLM suggests "Citrus" and "Camphorous", respectively. Here, we need to keep in mind that odor descriptors such as 'clean' are often culturally determined and may be linked to different odors in different countries. Overall, our comparison shows that the DT does not solely rely on conceptually related odor classes but includes the effect and quality that the descriptors share in the human world. In contrast, the ET classifies descriptors more on the odor impression let by pre-existing expectations from the side of the experts, their knowledge of chemistry, historical scent classifications and existing scent wheels used in perfumery[1]. This is also reflected in the selection of the grouping terms generated by the LLM, which tries to flatten the different descriptors within the DT classes back to a lower dimension that is similarly biased toward conceptual odor similarities as the expert derived classes.

### 3.3 Error analysis and Interpretability of the classification results using XGBoost

To shed some light into the underlying structure-odor relationships of our dataset and taxonomies, it is important to conduct an error analysis on misclassifications of the model as well as interpretability studies such as permutation feature importance (PFI). These insights help to verify that the model is learning meaningful patterns rather than artifacts, and allow identifying limitations and bottlenecks to guide future feature engineering and model refinements. The PFI plots of the XGBoost classifier (the best performing model see Section 3.1) is provided in Figure 5 for both taxonomies. Notably, several features originating from the Burden matrix such as "BCUTZ_1h" and "BCUTare-1l" exhibit significant drops in the AUC upon permutation, indicating their strong influence on the model performance. However, it should be noted that PFI is more indicative of overall performance of the model and that the class-specific relevance still needs to be taken into account to gain deeper insight into the underlying feature importance of the model.

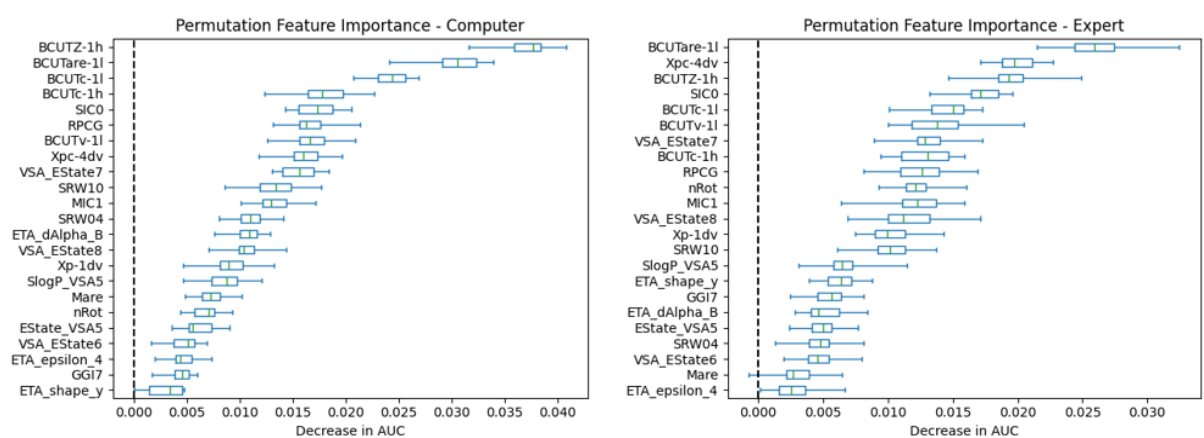

**Figure 6.** Permutation feature importance (PFI) of the XGBoost model for the data-driven taxonomy (left) and the expert-derived taxonomy (right). While such as BCUTZ-1h, BCUTare-1l and BCUTc-1l are very important to attain good macro scores with the data-driven taxonomy, the most relevant features for the expert taxonomy are BCUTZ-1h, BCUTare-1l and Xpc-4dv. Note that class-specific relevance needs to be taken into account to obtain more insight into the feature importance of the models.

Additionally, the SHAP plots for the Sulfur and Savory class can be seen in Figure 6 and Figure 7. The SHAP plots of all remaining classes for both taxonomies are provided in the ESI (FigureS14-S16). Altogether, the "Sulfur" and "Savory" classes are the best performing classes for both taxonomies with AUC over 0.75. The feature which contributes significantly to the Sulfur class across both taxonomies is 'BCUTZ-1h", which is particularly important within the SHAP values of the DT. The BCUTZ-1h feature represents the highest eigenvalue of the Burden matrix weighted by the atomic number. This corresponds to the atomic number of the heaviest atom in the compound, which is 16 in sulfur-containing compounds. This feature helps in classifying with both the "Sulfur" and "Savory" classes, which are both similar in odor profile (see Figure 7). There is also minor variation in the SHAP values of the Sulfur class across both taxonomies as well, where in the DT, the influence of BCUTZ-1h is more pronounced as compared to the ET where it shares similar weight with other features. This might be indicative of the odor descriptors under their respective umbrella terms, wherein the 'Sulfur' class in the DT is more narrow while that of the ET is a bit more general. We can see this in the distribution of BCUTZ-1h values across different taxonomies (see Figure S18): where in the DT, both the 'I' and 'J' classes, 'Savory' and 'Sulfur' respectively, have typically a higher BCUTZ-1h score with the Sulfur class being more pronounced, and in the ET the distribution of BCUTZ-1h is more similar for 'Savory', 'Smoky', and 'Sulfur.' It is interesting to note that there is a distinction between Savory and Sulfur and the features that help predict it. While BCUTZ-1h is the top feature for both taxonomies, when considering 'Savory', the feature SIC0 seems to be the most important predictor across both taxonomies. 'SIC0' is the structural information content of the molecule normalized by atom count. This feature corresponds to the degree of structural complexity where a high value suggests distinct atom types or diversity in atom degrees and a low value suggests more uniformity in atom types. The SHAP plots that for the Savory class, a higher SIC0 value pushes the model with towards a positive classification for Savory as compared to 'Sulfur' where interestingly, the inverse occur where for both taxonomies, a higher SIC0 value pushes the model against classifying Sulfur. This could imply that minute variations in distinct atom types or diverse atom degrees could explain the minor differences in the odor profiles of both "Sulfur" and "Savory".

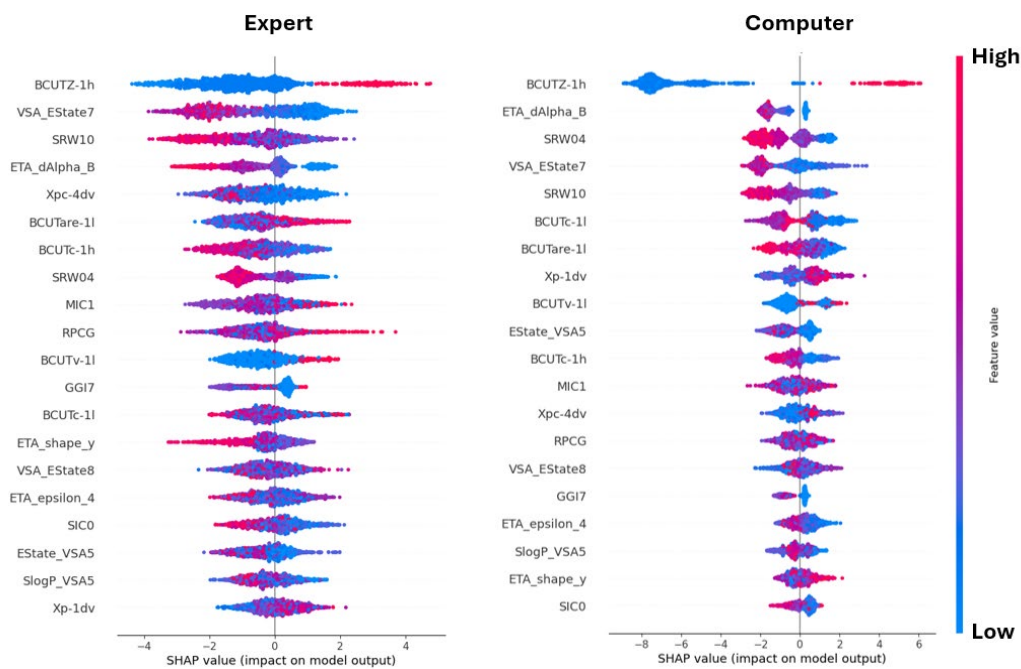

**Figure 7.** SHAP analysis of the XGBoost classifier class in both Taxonomies for the ET-defined "Sulfur" class and the corresponding DT class (Class "N"). We can see that the feature 'BCUTZ-1h' is listed as the most important feature for classifying 'Sulfur' across both taxonomies. For the full list of descriptors see supporting information.

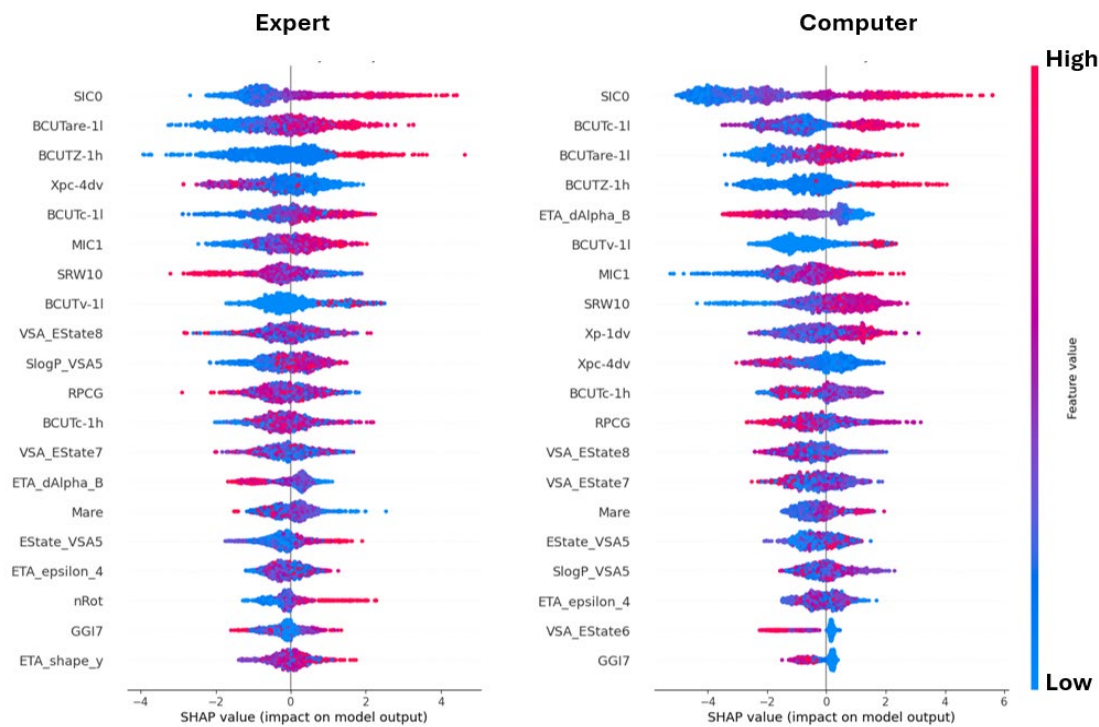

**Figure 8.** SHAP analysis of the XGBoost classifier class in both Taxonomies for the ET-defined "Savory" class and the corresponding DT class (class "I"). Note that in both taxonomies, the feature 'SIC0' has the biggest contribution, overtaking 'BCUTZ-1h' from the 'Sulfur' class for both taxonomies.

## 3.4 Testing the taxonomy models for class prediction: The case of pear odorants

To evaluate the applicability of the taxonomy models on interesting odorants used in perfumery, we selected five pear odorants and classified their odor description using the two XGBoost taxonomy based models. Pear odorants play a well-established role in perfumery due to their popularity as a top note in perfumes where they are often used to design fresh, bright, and juicy fragrances.

The molecular structures of the 4 pear odorant molecules used in perfumery, as well as their reported odor descriptors, are depicted in Figure 8. It should be noted that Pearadise (**2**) is an artificial odorant that was rationally designed based on known structures of pear odorants with the aim to be fully biodegradable and sustainable[33]. In future, next to specific odor and smell impressions, biodegradability and biocompatibility will become increasingly important and raise the challenge and requirements on machine learning methodologies within the fields of fragrance chemistry and in silico molecular design[34].

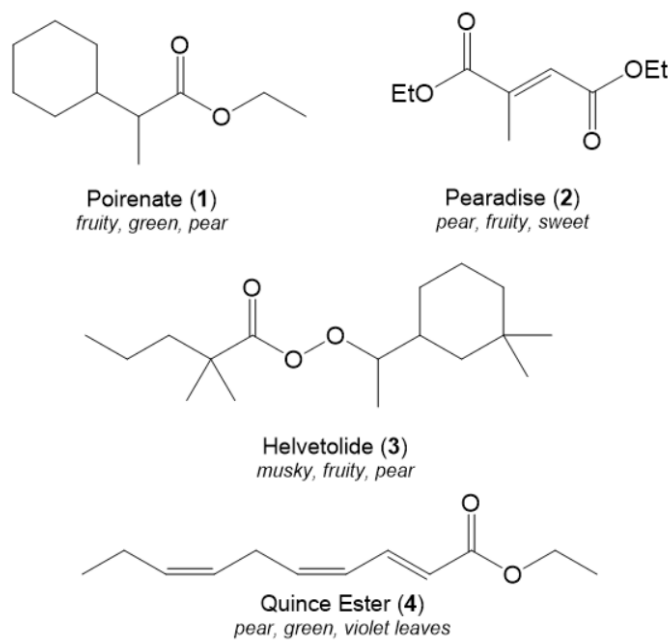

**Figure 9.** Selected examples of pear odorants. Note that all are correctly classified as 'fruity' using both taxonomies.

The selected subset of odorants serves as domain specific test cases to assess how well our models generalize to a broader training distribution. The selected type of odorants from this odor family are clearly related to the "Fruity" and the "J" classes of the ET and DT, respectively. The classifications results are shown in Table 4. Interestingly, both models are able to classify the most relevant odor descriptors of the pear odorants, namely "Fruity" and "Green". Going forward, it will be interesting to look into the explicit odor descriptors within the taxonomy groupings to distinguish more specific odors and continue to delineate the odor space and the structure-odor relationships linked to it.

**Table 4.** Classification results using the XGBoost model at the expert- and data-driven taxonomies for five selected pear odorant use for top notes in perfumery (see Figure 8 for the corresponding chemical structures).

| Pear Odorants | Data-driven Taxonomy[1] | Expert-derived Taxonomy |
|---|---|---|
| **Poirenate (1)** | C, J, O | Animal Body, Flower, Fruity, Woody |
| **Pearadise (2)** | J | Fruity |
| **Helvetolide (3)** | C, G, J | Fruity, Green, Woody |
| **Quinceester (4)** | G, J | Fruity, Green |

[1] Classes C, G, J, O can be conceptually mapped to the expert classes Flower, Green/Fat, Fruity, and Woody, respectively (see also Table 3 for a complete overview).

## 4 Outlook and perspectives

Currently, the most promising machine learning models for structure-based odor prediction are deep learning approaches, which have shown reliable performance in QSAR tasks[35,36]. However, conventional deep learning methods do not offer much in terms of interpretability. Therefore, our approach goes back to simpler models which offer more interpretability to aid insight into structure-odor relationships by leveraging the conceptual hierarchy between smell. Moving forward, graph-based architectures, particularly more recent ones like Fragnet[37] and KerGNNs[38], offer the potential for improved interpretability and more refined modeling of structure–activity relationships using deep learning. Odor taxonomies could be integrated into these GNNs not only to explore higher-level odor classes but also potentially within the loss function itself, guiding fine-grained classification across all descriptors. The ET, although explicitly flattened for this study, consists of multiple hierarchical layers that can be applied to design the GNN architecture. For the DT, the correlations between different odor descriptor groupings can also be further leveraged to create a multi-layer hierarchical taxonomy. Such methods could pave the way toward a deeper understanding of structure–odor relationships and the rational design of novel odorant molecules.

Despite the relevance of taxonomies for odor classification, there are several bottlenecks still limiting the progress that can be achieved in structure-odor prediction tasks. One important aspect is unarguably the 3D molecular structure of odorants. This work, as well as the current state-of-the-art rely on 2D molecular graphs as input, which omit conformational variability and spatial features that are likely relevant to odor perception. Odorants are inherently three-dimensional and can adopt multiple conformations, some of which may be more relevant to receptor binding and perceived odor than others. Incorporating 3D molecular geometry has already shown promise in related molecular classification tasks[39], and methods such as HamNet[40] that leverage molecular dynamics to represent conformations may extend this potential further. Still, receptor-level specificity adds another layer of complexity—olfactory receptors may respond differently to different conformers, but we currently lack sufficient experimental data to explore this[41].

In future, research that focuses on receptor-odorant dynamics and mechanisms could be leveraged to make better datasets and consequently, more well informed machine learning models. This would also allow us to account for differences between enantiomers and explore the impact of chirality[42]. Additionally, current datasets do not account for concentration-dependent effects and do not provide any information on the weight of odor descriptors, i.e., the dominant descriptors compared to less relevant side-notes. Concentration dependence is crucial, as the perceived odor of a molecule can vary with concentration. This effect is particularly strong for sulfur and nitrogen-containing compounds. To address these shortcomings, more high quality sensory data with different concentrations is required. Our work currently focuses on single odorant odor prediction without considering the effect of odorant mixture. In Olfaction however, smell sources emit hundreds of molecules that interact together to create a specific smell[43–45]. Therefore, to digitize smell and recreate odor compositions, it is necessary to move towards the study of odorant mixtures, concentrations, while keeping in mind that in some cases, olfactory groups and descriptors can additionally depend on social, religious, linguistic or other culturally induced categories[46,47].

## 5 Conclusion

In this work, we provide an openly accessible, well curated and structurally organized molecular odor dataset including two taxonomies (expert-derived and data-driven) that group odor descriptors in meaningful groupings that can be used to train machine learning models. We show that perceptual odor taxonomies can be incorporated into molecular structure–odor datasets as a form of data augmentation, allowing us to shift the classification task from high to low granularity, thereby improving the performance of machine learning tasks beyond a reduction in output space. A systematic benchmark using randomized groupings of odor descriptors shows that groupings without any conceptually meaningful connections do not improve the performance of machine learning models. In contrast, meaningful taxonomies based on grouping odors allow to address this challenge. Hereby, that data-driven taxonomy performs marginally better than the expert derived taxonomy. This study shows how multi-faceted odor taxonomies can be, from the comparison of our pre-conceived notions of distances between odor descriptors in the odor space according to expert taxonomy as well as the data taxonomy which takes into account the individual shared experiences with odor that shapes the topology of the odor space. Together with interpretable tree-based models this allowed us to link molecular structures to the respective odor space and pave the way for future machine learning approaches in structure-based odor prediction. Our framework thus lays the foundation for more accurate, interpretable, and scalable approaches to decoding the language of smell, and help us move closer towards the digitization of one of our most enigmatic senses.

## 6. Data availability

The MMD, the taxonomies (ET and DT) including their taxonomy augmented datasets, and annotated code, for all the methodology as well as the analyses, are included as Supplementary material and will be publicly available on [GitHub](GitHub).

**Acknowledgements**

This publication is part of the project ODORWISE with file number OCENW.M.23.020 of the research programme Open Competition Domain Science M23-1 which is (partly) financed by the Dutch Research Council (NWO). The authors acknowledge the SURFsara compute cluster hosted by SURF and the BAZIS research cluster hosted by VU for the computational time and the provided technical support.

**Author contributions statement**

All authors, AS, SA, RT, PL, IL, SS, CV, RH, HM were involved in the work. AS performed the data analyses and interpretation; AS and H.M. prepared the original draft of the manuscript; HM, RH provided supervision for AS and SS, SS and AS carried out the initial literature search and overview; HM, SA conceptualized the project idea, HM was responsible for funding and main supervision. IL, CV, PL, RT, HM derived the expert-based taxonomy, AS and SS provided the clustering, AS implemented the randomized grouping and provided the Github repository. PL and RT structured the expert taxonomy and made it available on the ODEUROPA website, and offered feedback about the technical methodology and development to enhance the work. All authors critically reviewed, edited the manuscript, and approved the final version of the manuscript.

**Competing interests.** The authors declare no competing interests.

# Electronic Supporting information

# Exploring molecular odor taxonomies for structure-based odor predictions using machine learning

Akshay Sajan,[a] Stijn Sluis,[a] Reza Haydarlou,[a] Sanne Abeln,[b] Pasquale Lisena,[c] Raphael Troncy,[c] Caro Verbeek,[d] Inger Leemans,[e] and Halima Mouhib[a]*

[a] Department of Computer Science, VU Bioinformatics Group, Vrije Universiteit Amsterdam, De Boelelaan 1105, 1081 HV Amsterdam, The Netherlands.
[b] Department of Computer Science, AI Technology for Life, Universiteit Utrecht, Heidelberglaan 8, 3584 CS Utrecht, The Netherlands
[c] EURECOM, Campus Sophia Tech, 450 Route des Chappes, 06410 Biot, France.
[d] Faculty of Humanities, Art and Culture, History, Antiquity, De Boelelaan 1105, 1081 HV Amsterdam, The Netherlands
[e] KNAW Humanities Cluster, Oudezijds Achterburgwal 185, 1012 DK Amsterdam, The Netherlands. Vrije Universiteit Amsterdam, Department of Art and culture, History, and Antiquity, Faculty of Social Sciences and Humanities, De Boelelaan 1105, 1081 HV Amsterdam, The Netherlands.

*corresponding author: h.mouhib@vu.nl

**Molecular odor descriptors for the provided molecular data set used to build the machine learning models (in alphabetical order).**

['acid', 'alcohol', 'aldehyde', 'almond', 'amber', 'animal', 'anise', 'anisic', 'apple', 'apricot', 'balsam', 'banana', 'beef', 'bergamot', 'berry', 'bitter', 'black currant', 'brandy', 'bread', 'broth', 'burnt', 'butter', 'cabbage', 'camphor', 'caramel', 'cedar', 'celery', 'chamomile', 'cheese', 'chemical', 'cherry', 'chicken', 'chocolate', 'cinnamon', 'citrus', 'clove', 'cocoa', 'coconut', 'coffee', 'cognac', 'cooked', 'coumarin', 'cream', 'cucumber', 'dairy', 'earth', 'ester', 'ether', 'fat', 'fermented', 'fish', 'floral', 'flower', 'fruity', 'gardenia', 'garlic', 'gasoline', 'gassy', 'geranium', 'gourmand', 'grape', 'grapefruit', 'grass', 'green', 'hawthorn', 'hay', 'hazelnut', 'herbal', 'honey', 'horseradish', 'hyacinth', 'jam', 'jasmine', 'juice', 'ketonic', 'lactonic', 'lavender', 'leaf', 'lemon', 'licorice', 'lilac', 'lily', 'malt', 'marine', 'meat', 'medicinal', 'melon', 'menthol', 'metallic', 'milk', 'mimosa', 'mint', 'moss', 'muguet', 'mushroom', 'musk', 'musty', 'narcissus', 'neroli', 'onion', 'orange', 'orris', 'ozone', 'patchouli', 'peach', 'pear', 'peel', 'pepper', 'phenol', 'pine', 'pineapple', 'plastic', 'plum', 'popcorn', 'potato', 'pungent', 'raspberry', 'ripe', 'roasted', 'rooty', 'rose', 'rum', 'sandalwood', 'savory', 'sharp', 'smoked', 'solvent', 'sour', 'spicy', 'strawberry', 'sulfur', 'sweat', 'tea', 'terpene', 'tobacco', 'tomato', 'tropical', 'vanilla', 'vegetable', 'vetiver', 'violet', 'watery', 'wax', 'weedy', 'wine', 'woody']- **total of 146 descriptors**

**Feature selection and Hyperparameter tuning**

The features were filtered through an exhaustive feature selection procedure. 0 variance features were removed and then ANOVA F values were used to filter the features and the best feature for each of the 146 classes was selected. This was then narrowed down further through Recursive Feature Elimination, using a random forest with Permutation Feature Importance. The final set contained 23 features. The hyperparameter tuning of the models was done using Bayesian optimization through skopt [HEA].

**Parameters for clustering odor descriptors in Computer Taxonomy**
[n_clusters = 16, metric = 'euclidean', linkage ='ward']

**Final features:**

['BCUTc-1h', 'BCUTc-1l', 'BCUTZ-1h', 'BCUTv-1l', 'BCUTare-1l', 'RPCG', 'Xpc-4dv', 'Xp-1dv', 'Mare', 'ETA_shape_y', 'ETA_dAlpha_B', 'ETA_epsilon_4', 'SIC0', 'MIC1', 'SlogP_VSA5', 'EState_VSA5', 'VSA_EState6', 'VSA_EState7', 'VSA_EState8', 'nRot', 'GGI7', 'SRW04', 'SRW10']

**Table S1.** Search space and final parameters of the hyperparameter search for the XGBoost model using Bayesian Hyperparameter Optimization.

| Hyperparameter | Search Space | Chosen Value |
|:---:|:---:|:---:|
| colsample_bynode | [0.1, 0.2, 0.3, 0.4, 0.5, 0.6, 0.7, 0.8, 0.9, 1.0] | 0.9 |
| learning_rate | [0.0001, 0.001, 0.01, 0.1, 0.2, 0.4, 0.6, 0.8] | 0.4 |
| max_depth | [3, 4, 5, 6, 7, 8, 9, 10, 11, 12] | 3 |
| min_child_weight | [1, 50, 100, 150, 200, 250] | 1 |
| n_estimators | [100, 200, 400, 600, 800, 1000, 2000, 4000, 5000, 10000] | 2000 |
| reg_lambda | [0.001, 0.01, 0.1, 1, 5, 10, 15, 20, 25] | 5 |
| subsample | [0.1, 0.2, 0.3, 0.4, 0.5, 0.6, 0.7, 0.8, 0.9, 1.0] | 0.9 |
| tree_method | ['approx', 'hist'] | 'hist' |

**Table S2.** Search space and final parameters of the hyperparameter search for the Random Forest model using Bayesian Hyperparameter Optimization.

| Hyperparameter | Search Space | Chosen Value |
|---|---|---|
| bootstrap | [True, False] | 'False' |
| max_depth | [10, 20, 30, 40, 50, 60, 70, 80, 90, 100, None] | 90 |
| max_features | ['log2', 'sqrt'] | 'log2' |
| min_samples_leaf | [1, 2, 3, 4] | 1 |
| min_samples_split | [2, 5, 10] | 2 |
| n_estimators | [50, 150, 200, 400, 600, 800, 1000, 1200, 1400, 1600, 1800, 2000] | 1800 |

**Table S3.** Search space and final parameters of the hyperparameter search for the multi output Logistic Regression model using Bayesian Hyperparameter Optimization.

| Hyperparameter | Search Space | Chosen Value |
|---|---|---|
| penalty | ['l2'] | 'l2' |
| C | np.logspace(-4, 4, 20) | 10000.0 |
| solver | ['lbfgs','newton-cg','liblinear','sag','saga'] | 'lbfgs' |
| max_iter | [100, 1000, 2500, 5000] | 2500 |

**Table S4.** Performance results using a **Logistic Regression model** after extensive feature selection and hyperparameter tuning.

| Dataset | AUROC | F1 Score | Precision | Recall |
|---|---|---|---|---|
| Original Descriptors | 0.527 | 0.079 | 0.170 | 0.059 |
| Expert Taxonomy | 0.603 | 0.334 | 0.547 | 0.260 |
| Computer Taxonomy | 0.609 | 0.319 | 0.528 | 0.262 |

**Table S5.** Performance results using a **Random Forest model** after extensive feature selection and hyperparameter tuning.

| Dataset | AUROC | F1 Score | Precision | Recall |
|---|---|---|---|---|
| Original Descriptors | 0.589 | 0.243 | 0.408 | 0.186 |
| Expert Taxonomy | 0.676 | 0.488 | 0.601 | 0.421 |
| Computer Taxonomy | **0.693** | **0.511** | **0.623** | **0.444** |

**Table S6.** Classwise performance of the XGBoost model using the computer derived taxonomy. The 16 classes (A-P) as well as their scores across all metrics. The macro (unweighted) average of all the classes is provided in the last row.

| Classes | AUROC | F1 | Precision | Recall |
|---|---|---|---|---|
| A | 0.659429 | 0.404762 | 0.485714 | 0.346939 |
| B | 0.674819 | 0.486275 | 0.529915 | 0.449275 |
| C | 0.726494 | 0.647120 | 0.674672 | 0.621730 |
| D | 0.654029 | 0.422330 | 0.446154 | 0.400922 |
| E | 0.702355 | 0.526971 | 0.569507 | 0.490348 |
| F | 0.664019 | 0.421569 | 0.505882 | 0.361345 |
| G | 0.717981 | 0.646245 | 0.670082 | 0.624046 |
| H | 0.708859 | 0.515406 | 0.557576 | 0.479167 |
| I | 0.756446 | 0.606335 | 0.697917 | 0.536000 |
| J | 0.720628 | 0.654369 | 0.693416 | 0.619485 |
| K | 0.699158 | 0.503704 | 0.596491 | 0.435897 |
| L | 0.708253 | 0.563107 | 0.610526 | 0.522523 |
| M | 0.641302 | 0.355140 | 0.431818 | 0.301587 |
| N | 0.885479 | 0.750000 | 0.709091 | 0.795918 |
| O | 0.696117 | 0.526733 | 0.588496 | 0.476703 |
| P | 0.558067 | 0.183908 | 0.320000 | 0.129032 |
| **Average** | **0.698340** | **0.513373** | **0.567954** | **0.474432** |

**Table S7.** Classwise performance of the XGBoost model using the expert derived taxonomy. The 16 classes as well as their scores across all metrics. The macro (unweighted) average of all the classes is provided in the last row.

| Classes | AUROC | F1 | Precision | Recall |
|---|---|---|---|---|
| Alcohol | 0.697169 | 0.502283 | 0.526316 | 0.480349 |
| Animal Body | 0.668494 | 0.433566 | 0.553571 | 0.356322 |
| Aquatic | 0.587067 | 0.256410 | 0.454545 | 0.178571 |
| Balsamic | 0.685158 | 0.482385 | 0.542683 | 0.434146 |
| Chemicals | 0.644282 | 0.419913 | 0.451163 | 0.392713 |
| Earthy | 0.592802 | 0.284615 | 0.342593 | 0.243421 |
| Flower | 0.748530 | 0.650246 | 0.668354 | 0.633094 |
| Fruity | 0.738040 | 0.715421 | 0.748752 | 0.684932 |
| Gourmand | 0.674920 | 0.483940 | 0.548544 | 0.432950 |
| Green | 0.681113 | 0.631034 | 0.640981 | 0.621392 |
| Herbal | 0.668194 | 0.424581 | 0.506667 | 0.365385 |
| Savory | 0.763273 | 0.664935 | 0.684492 | 0.646465 |
| Smoky | 0.692512 | 0.472441 | 0.526316 | 0.428571 |
| Spices | 0.647363 | 0.397260 | 0.456693 | 0.351515 |
| Sulfur | 0.761012 | 0.588710 | 0.623932 | 0.557252 |
| Woody | 0.696945 | 0.528958 | 0.585470 | 0.482394 |
| **Average** | **0.684180** | **0.496044** | **0.553817** | **0.455592** |

**Table S8. Odor descriptors falling under the 16 expert derived classes.**

| Class | # descriptors | List of descriptors |
|---|---|---|
| Alcohol | 10 | ['acid', 'sharp', 'pungent', 'brandy', 'cognac', 'ether', 'malt', 'rum', 'wine', 'alcohol'] |
| Animal Body | 4 | ['amber', 'musk', 'sweat', 'animal'] |
| Aquatic | 3 | ['fish', 'marine', 'watery'] |
| Balsamic | 2 | ['balsam', 'wax'] |
| Chemicals | 12 | ['ether', 'ozone', 'aldehyde', 'gasoline', 'ketonic', 'medicinal', 'metallic', 'phenol', 'plastic', 'solvent', 'terpene', 'chemical'] |
| Earthy | 5 | ['moss', 'mushroom', 'musty', 'earth', 'rooty'] |
| Flower | 15 | ['geranium', 'lavender', 'mimosa', 'narcissus', 'orris', 'rose', 'violet', 'gardenia', 'hyacinth', 'jasmine', 'lilac', 'lily', 'muguet', 'flower', 'floral'] |
| Fruity | 28 | ['berry', 'cherry', 'black currant', 'raspberry', 'strawberry', 'bergamot', 'citrus', 'grapefruit', 'lemon', 'neroli', 'orange', 'peel', 'apple', 'apricot', 'banana', 'coconut', 'ester', 'grape', 'hawthorn', 'juice', 'melon', 'peach', 'pear', 'pineapple', 'plum', ['ripe','tropical', 'fruity'] |
| Gourmand | 13 | 'almond', 'roasted', 'bitter', 'caramel', 'cocoa', 'cream', 'hazelnut', 'honey', 'jam', 'popcorn', 'vanilla', 'chocolate', 'gourmand'] |
| Green | 9 | ['celery', 'cucumber', 'grass', 'green', 'leaf', 'herbal', 'weedy', 'coumarin', 'hay'] |
| Herbal | 5 | ['menthol', 'mint', 'chamomile', 'tea', 'vetiver'] |
| Savory | 23 | ['garlic', 'onion', 'cooked', 'cabbage', 'roasted', 'bread', 'butter', 'cheese', 'cream', 'milk', 'lactonic', 'sour', 'fat', 'peel', 'beef', 'chicken', 'meat', 'broth', 'vegetable', 'potato', 'tomato', 'dairy', 'savory'] |
| Smoky | 5 | ['burnt', 'smoked', 'coffee', 'roasted', 'tobacco'] |
| Spices | 8 | ['anisic', 'cinnamon', 'clove', 'horseradish', 'licorice', 'pepper', 'anise', 'spicy'] |
| Sulfur | 3 | ['fermented', 'gassy', 'sulfur'] |
| Woody | 9 | ['camphor', 'cognac', 'patchouli', 'rooty', 'sandalwood', 'woody', 'lactonic', 'cedar', 'pine'] |

As described in Section 2.2.1 of the ms. for the expert taxonomy on the [ODEUROPA website](), the source-based descriptors were divided in 16 different classes 'scent families') and 31 Subclasses, e.g., the "Alcohol" class contains the sub-classes "Acid" and "Alcohol", which each contains their own odor descriptors. The classes (**bold**) and sub-classes (*italic*) are provided below, the hierarchical expert taxonomy is provided on the ODEUROPA website.

1. **Alcohol**
   1.1. *Acid*
   1.2. *Alcohol*
2. **Animal**
   2.1. *Animal*
   2.2. *Body*
3. **Aquatic**
   3.1. *Fish*
   3.2. *Sea*
4. **Balsamic**
5. **Chemical**
   5.1. *Ether*
   5.2. *Other*
6. **Earthy**
   6.1. *White flowers*
   6.2. *Other*
7. **Flower**
   7.1. *Berry*
   7.2. *Citrus*
   7.3. *Other*
8. **Fruity**
   8.1. *Ether*
   8.2. *Other*
9. **Gourmand**
10. **Green**
    10.1. *Grass*
    10.2. *Hay*
11. **Herbal**
    11.1. *Methol*
    11.2. *Other*
12. **Savory**
    12.1. *Allium*
    12.2. *Brassica*
    12.3. *Bread*
    12.4. *Dairy*
    12.5. *Fat*
    12.6. *Meat*
    12.7. *Umami*
    12.8. *Other*
13. **Smoky**
14. **Spices**
15. **Sulfur**
    15.1. *Decay*
    15.2. *Excrement*
    15.3. *Sulfur*
16. **Woody**
    16.1. *Ether*
    16.2. *Other*

**Table S9. Descriptors falling under the 16 computer derived classes.**

| Class | # descriptors | List of descriptors |
|---|---|---|
| A | 12 | ['almond', 'anise', 'anisic', 'bitter', 'cherry', 'hawthorn', 'hyacinth', 'licorice', 'lilac','mimosa', 'narcissus', 'plastic'] |
| B | 14 | ['bread', 'burnt', 'caramel', 'chocolate', 'cocoa', 'coffee', 'earth', 'gourmand', 'hazelnut', 'malt', 'mushroom', 'musty', 'popcorn', 'potato'] |
| C | 22 | ['bergamot', 'black currant', 'celery', 'chamomile', 'floral', 'flower', 'gardenia', 'geranium', 'grape', 'grapefruit', 'honey', 'jasmine', 'lavender', 'lily', 'muguet', 'neroli','orris', 'plum', 'rose', 'tea', 'tobacco', 'violet'] |
| D | 13 | ['cabbage', 'chemical', 'fish', 'gasoline', 'gassy', 'horseradish', 'ketonic', 'metallic', 'pepper', 'pungent', 'sharp', 'tomato', 'vegetable'] |
| E | 8 | ['balsam', 'cinnamon', 'clove', 'medicinal', 'phenol', 'smoked', 'spicy', 'vanilla'] |
| F | 8 | ['butter', 'coconut', 'coumarin', 'cream', 'dairy', 'hay', 'lactonic', 'milk'] |
| G | 8 | ['cucumber', 'fat', 'grass', 'green', 'leaf', 'melon', 'wax', 'weedy'] |
| H | 8 | ['alcohol', 'brandy', 'cognac', 'ether', 'fermented', 'rum', 'solvent', 'wine'] |
| I | 7 | ['beef', 'broth', 'chicken', 'cooked', 'meat', 'roasted', 'savory'] |
| J | 11 | ['apple', 'apricot', 'banana', 'ester', 'fruity', 'juice', 'peach', 'pear', 'pineapple', 'ripe', 'tropical'] |
| K | 8 | ['aldehyde', 'citrus', 'lemon', 'marine', 'orange', 'ozone', 'peel', 'watery'] |
| L | 6 | ['camphor', 'herbal', 'menthol', 'mint', 'pine', 'terpene'] |
| M | 4 | ['acid', 'cheese', 'sour', 'sweat'] |
| N | 3 | ['garlic', 'onion', 'sulfur'] |
| O | 10 | ['amber', 'animal', 'cedar', 'moss', 'musk', 'patchouli', 'rooty', 'sandalwood', 'vetiver', 'woody'] |
| P | 4 | ['berry', 'jam', 'raspberry', 'strawberry'] |

**The prompt used to generate the class names with ChatGPT (GPT-4o) is the following:**
"I am working on a taxonomy (or ontology) to describe the conceptual hierarchy between odor descriptors for molecules. This will be leveraged to improve machine learning models for structure-odor predictions. I have groups of several descriptors, e.g., "rose", "meat", "blackcurrant" etc and I want to give a name to each group, an umbrella term under which different descriptors can fall. The descriptors' groups are the following: [...]
I need only a single word to best describe each group, with vocabulary that is mainly associated with perfumery and smell experts. Please go in the right order that I provide."

**See Section 3.2 of the ms. and Table 3 for the list of class names.**

**Randomization convergence.** The total number of possible randomized taxonomies that can be generated using the odor descriptors in their respective parent term is 146 factorial (146!). However, to get an overview of the score distribution of each metric it is not necessary to use all the 146! possible combinations. According to the central limit theorem, the sample distribution converges to the population distribution with increasing sample size. It is therefore possible to approximate the population mean with a lower number of samples. To verify if the running mean and running standard deviation for all metrics have converged, we use 1000 randomized taxonomies. The randomization convergence plots of the expert taxonomy (1000), computer taxonomy (1000) and the combined taxonomies (2000) are shown below.

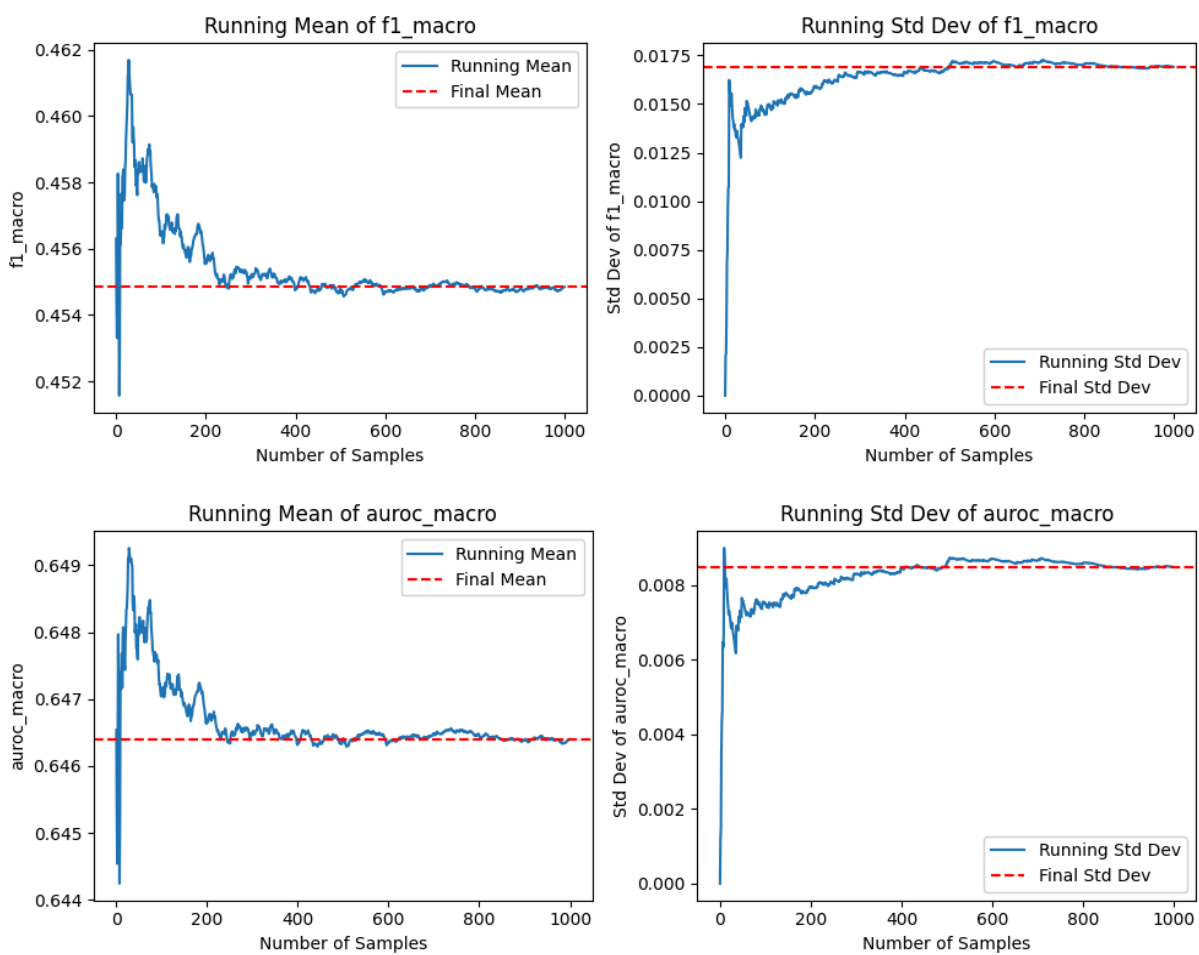

**Figure S1-1. Development of the mean and standard deviation with the addition of each new metric score for 1000 randomizations on the expert derived taxonomy.** Macro metrics F1 and AUROC over number of samples (randomized taxonomies). Note that all metrics converge, showing that the score metrics can be taken as a close representative of the randomizations, or the gain in score metrics due to reducing the number of classes.

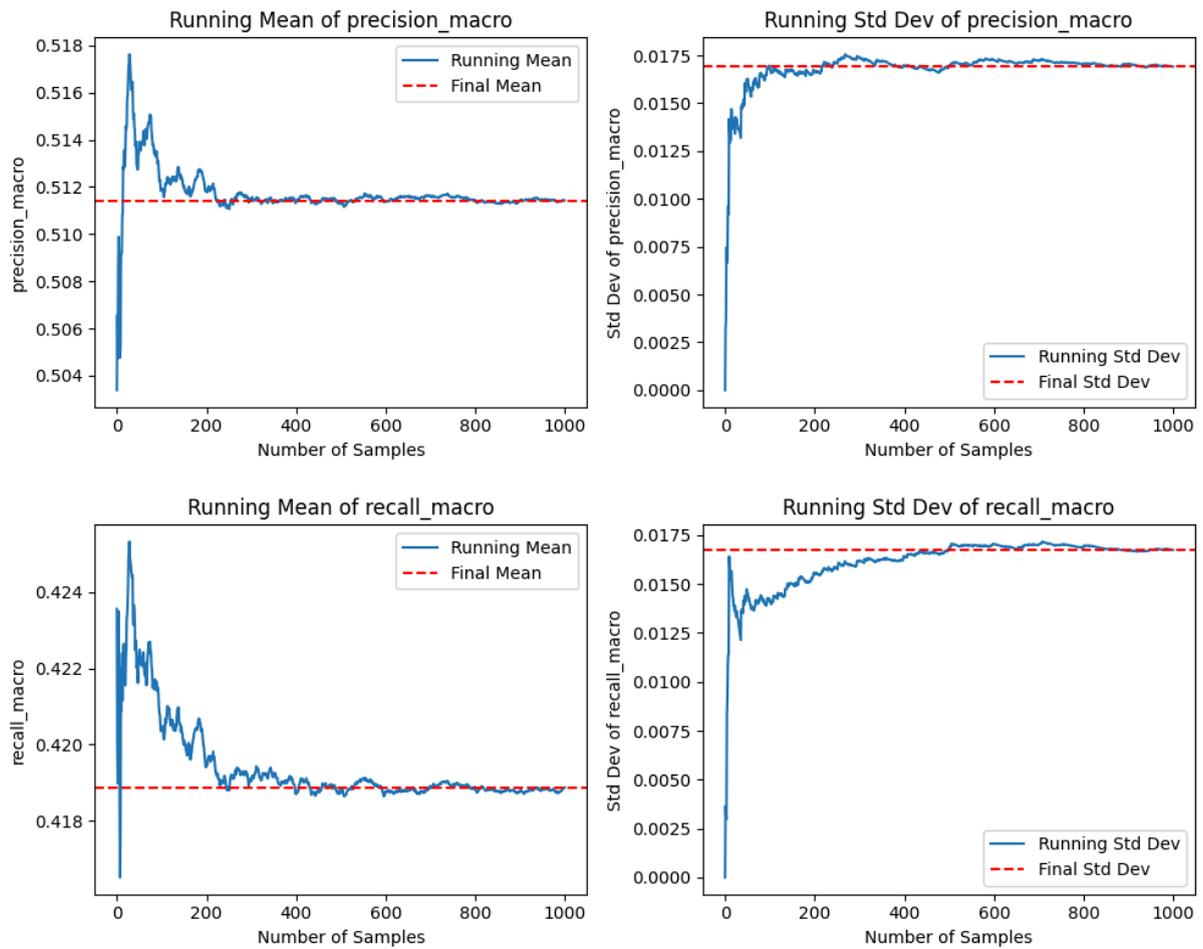

**Figure S1-2.** Macro metrics precision and recall (randomization expert derived taxonomy). See Caption Figure S1-1 for detailed description.

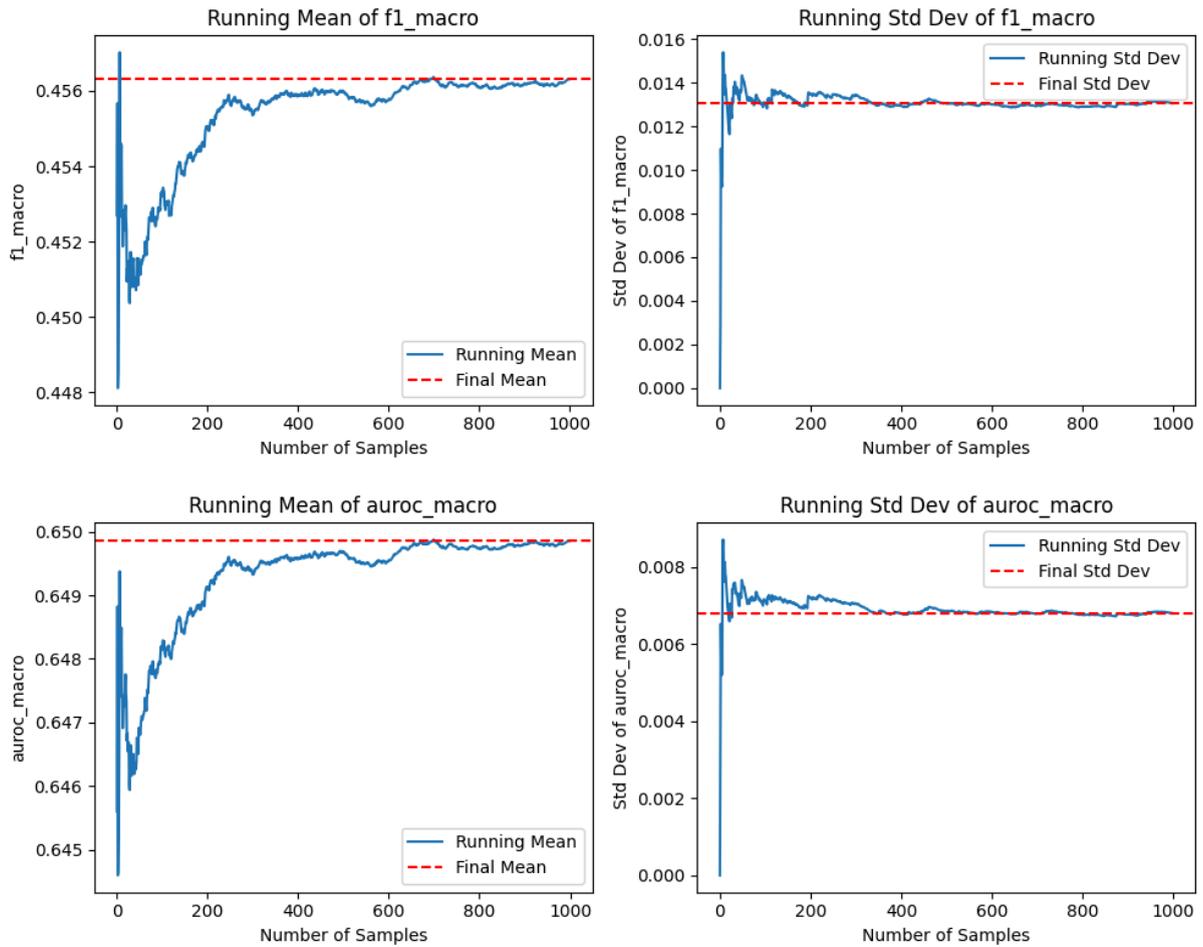

**Figure S2-1. Development of the mean and standard deviation with the addition of each new metric score for 1000 randomizations on the computer derived taxonomy.** Macro metrics F1 and AUROC over number of samples (randomized taxonomies). Note that all metrics converge, showing that the score metrics can be taken as a close representative of the randomizations, or the gain in score metrics due to reducing the number of classes.

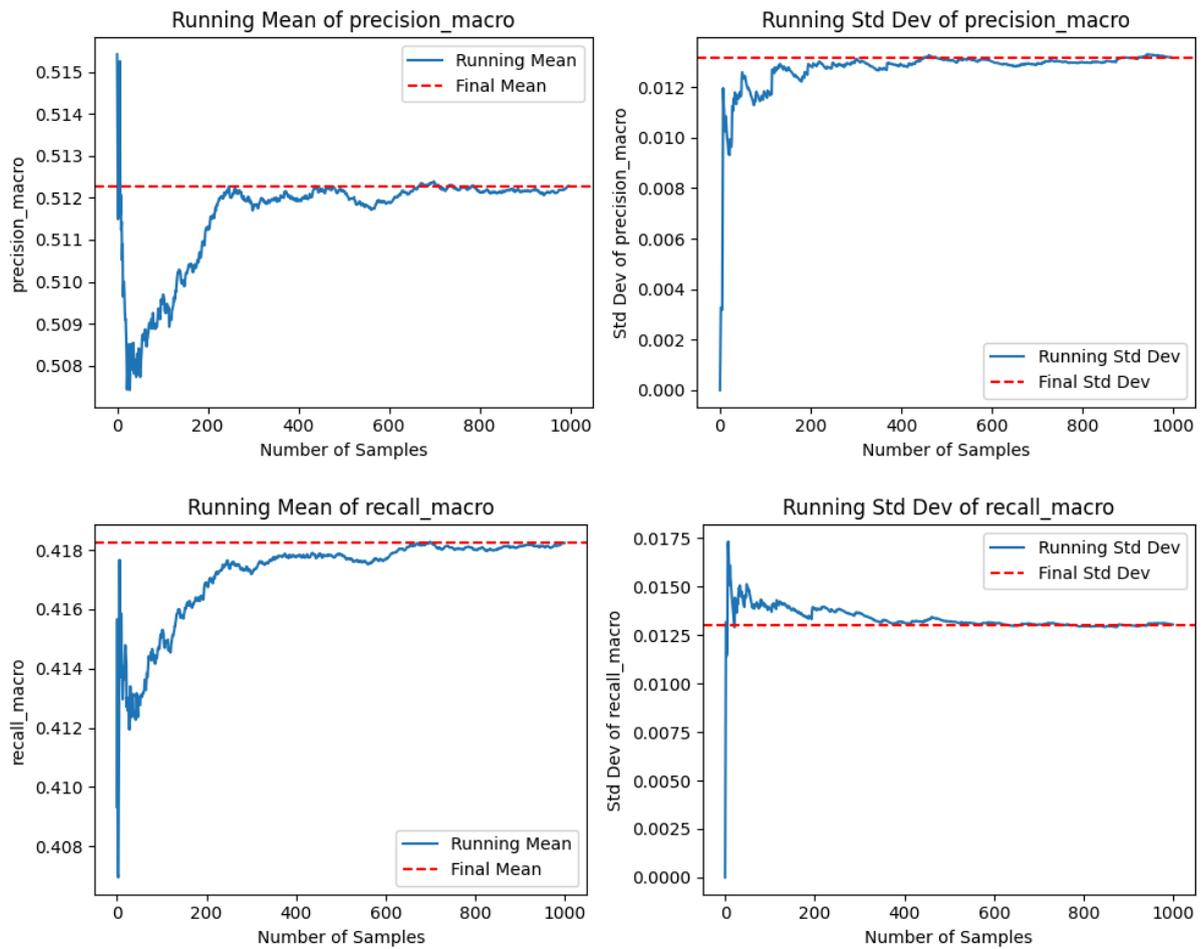

**Figure S2-2.** Macro metrics precision and recall (randomization computer derived taxonomy). See Caption Figure S2-1 for detailed description.

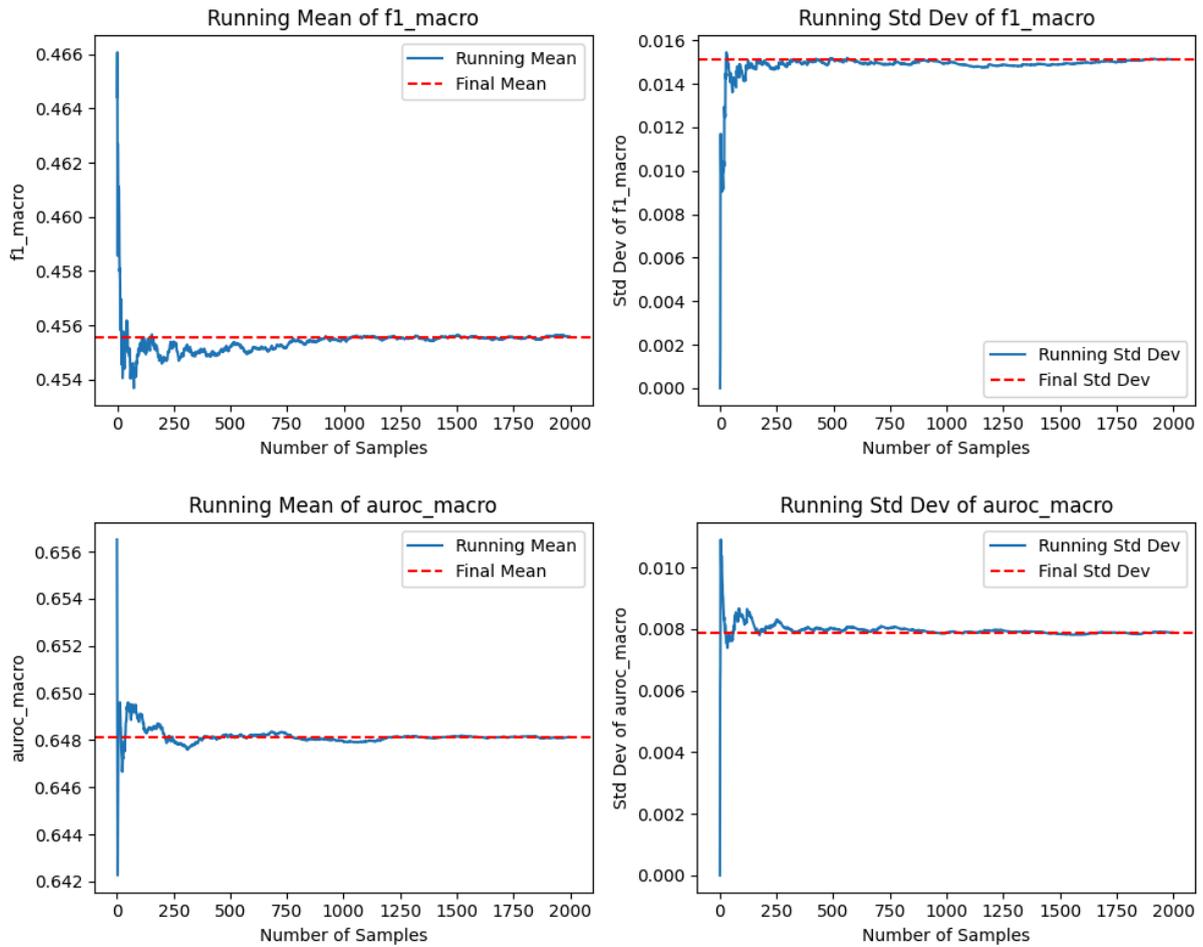

**Figure S3-1.** Mean and standard deviation over 2000 randomizations taxonomies combining the macro metrics F1 and AUROC for both the expert and the computer derived taxonomy (see Figures S1 and S2, respectively). This shows that the joint distribution converges and can be used to represent the performance gain from reducing class count.

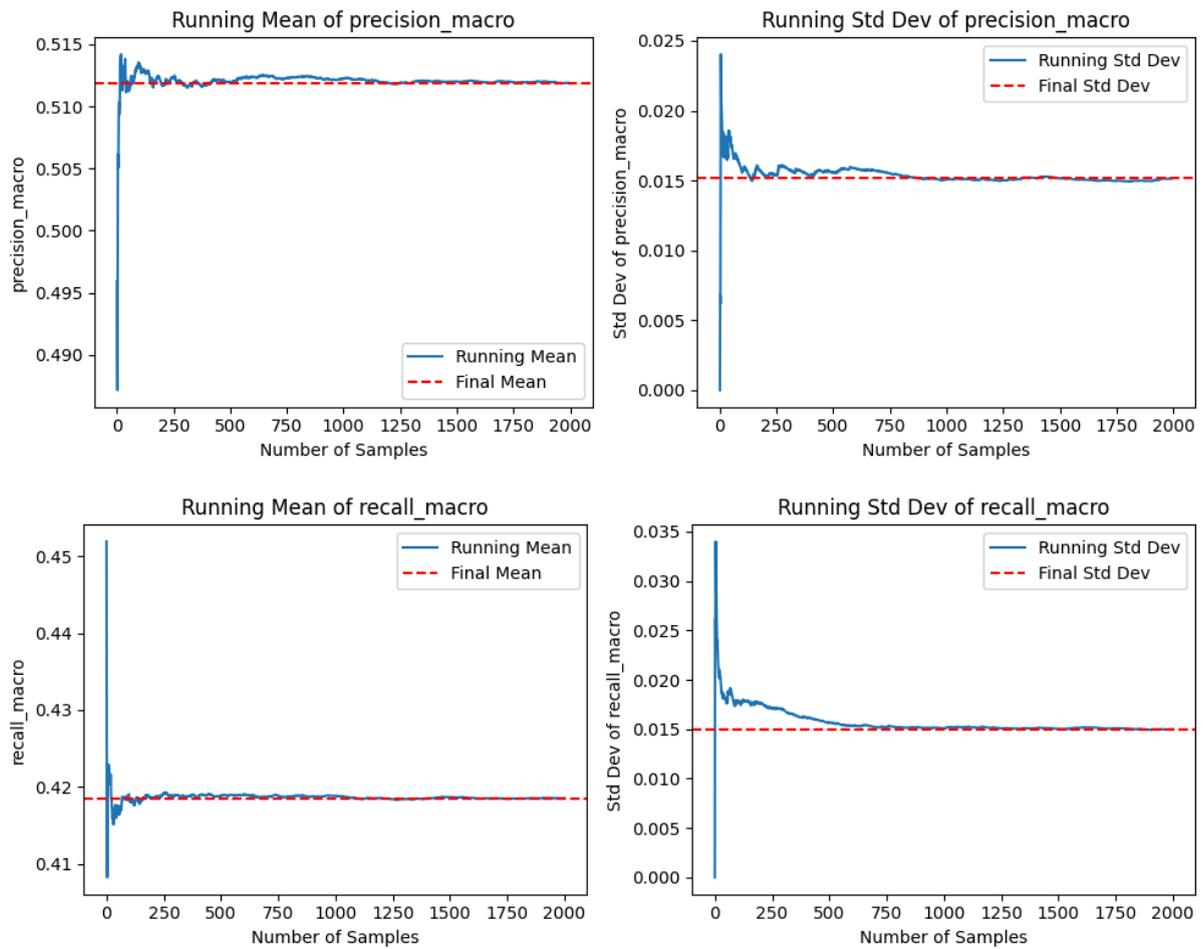

**Figure S3-2.** Mean and standard deviation over 2000 randomizations taxonomies combining the macro metrics precision and recall for both the expert and the computer derived taxonomy (see Figures S1 and S13, respectively). See Caption Figure S3-1 for detailed description.

**Data Exploration**

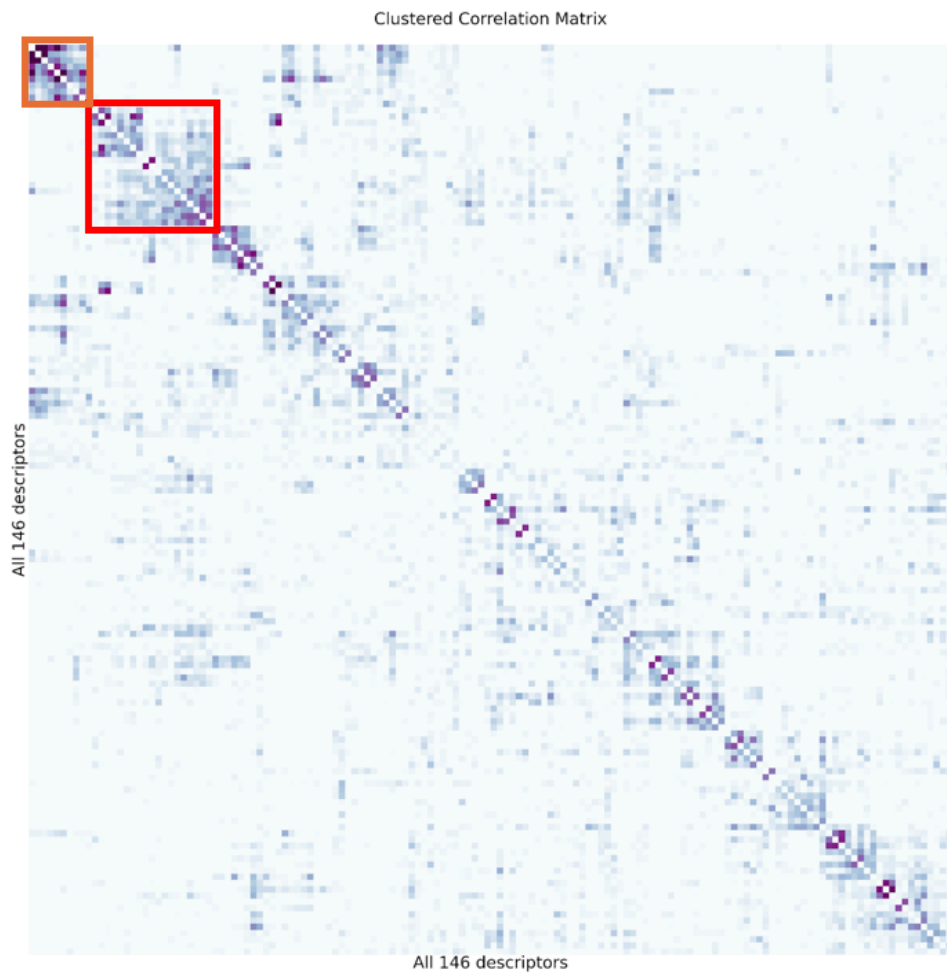

**Figure S4.** Clustered correlation plot of all 146 descriptors using the co-occurrence from the training dataset. The co-occurrence of similar odor descriptors allows for a more perceptually meaningful clustering. The two clusters highlighted in orange and red correspond to the sulfur/savory and the alcohol/fruity clusters, respectively. For a zoom in on the alcohol and the fruity clusters see manuscript (Figure 4).

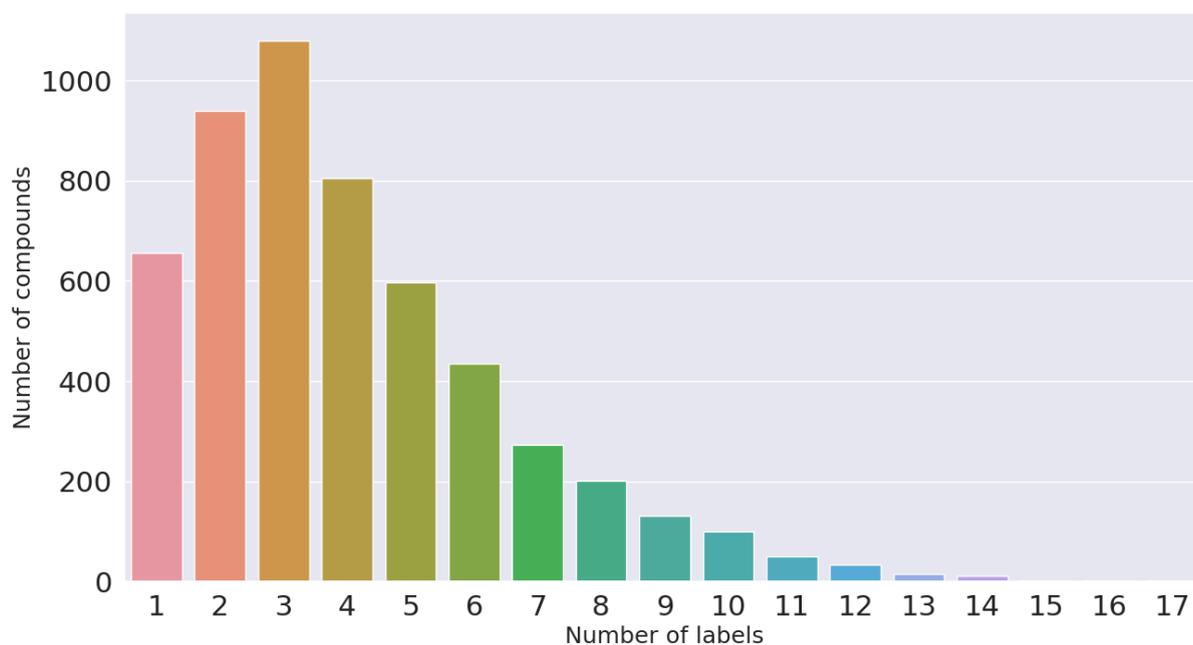

**Figure S5.** Overview of chemical compounds and the number of corresponding labels or descriptors provided in the full training dataset (5331 molecules - see ms. for GitHub repository). The majority of compounds have between 1 and 5 labels. 32 compounds have 13 labels or more with a maximum number of 17 descriptors.

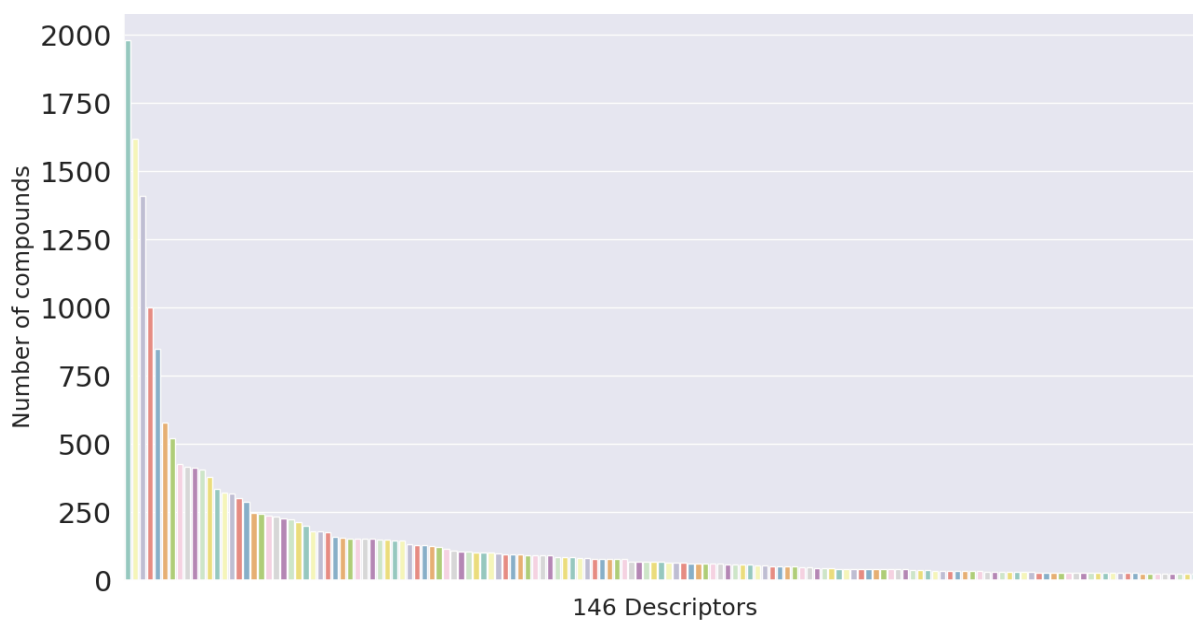

**Figure S6.** Number of compounds in each of the different classes, i.e., provided smell descriptors, for a total of 146 descriptors. It should be noted that the number of instances throughout the classes are highly imbalanced.

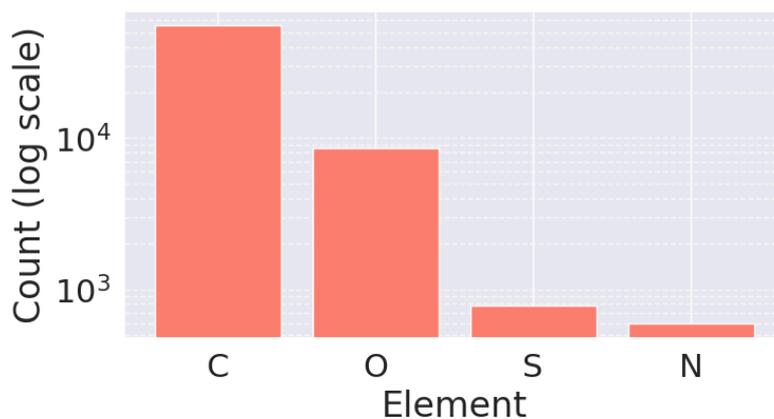

**Figure S7.** Total sum of the number of carbon (C), oxygen (O), sulphur (S), and nitrogen (N) atoms within the all-descriptor training dataset (5331 molecules).

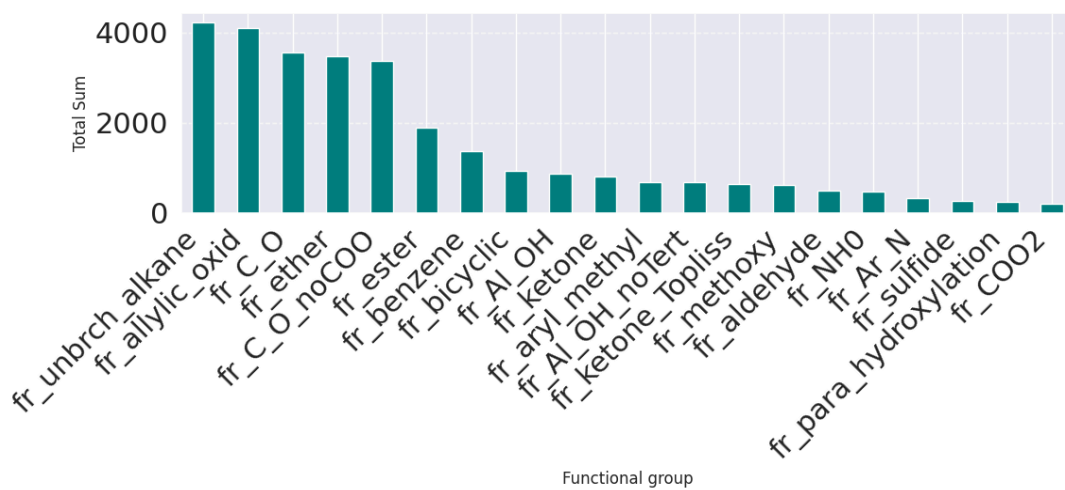

**Figure S8.** Number of different functional groups present in the training data set (5331 compounds - 80% of the full dataset with 6711 molecular compounds).

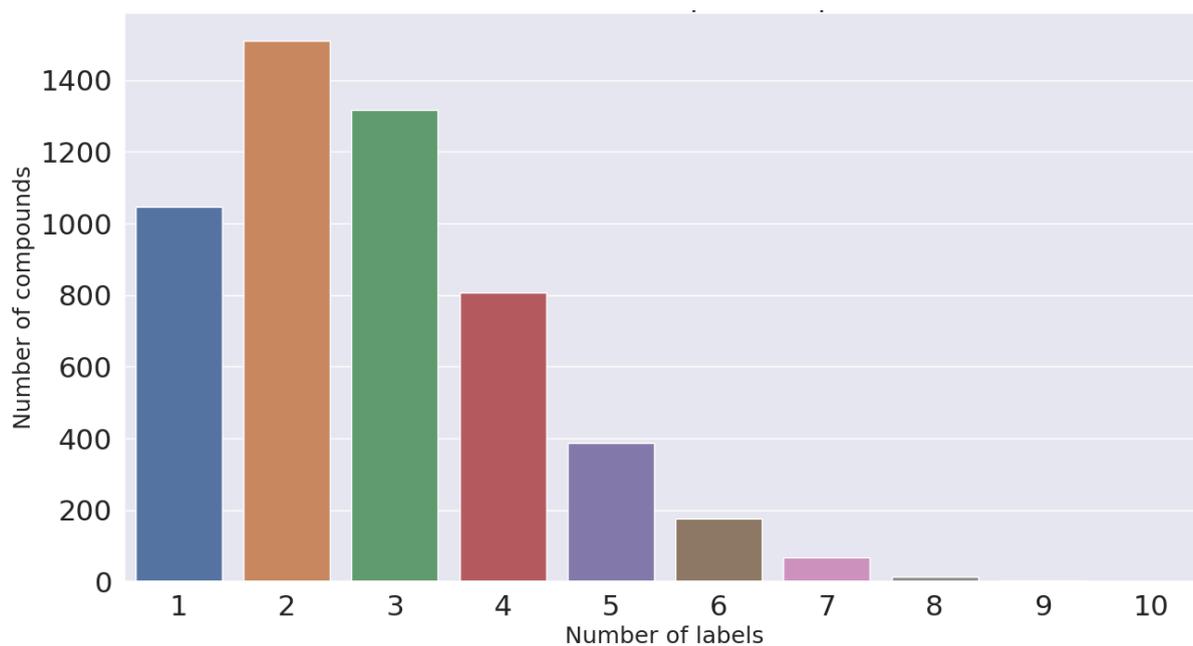

**Figure S9.** Overview of chemical compounds and the number of corresponding labels or descriptors provided in the data set imposed using the computer derived taxonomy. The majority of compounds have between 1 and 4 labels. See Figure S8 for the distribution of the labels obtained using the expert taxonomy.

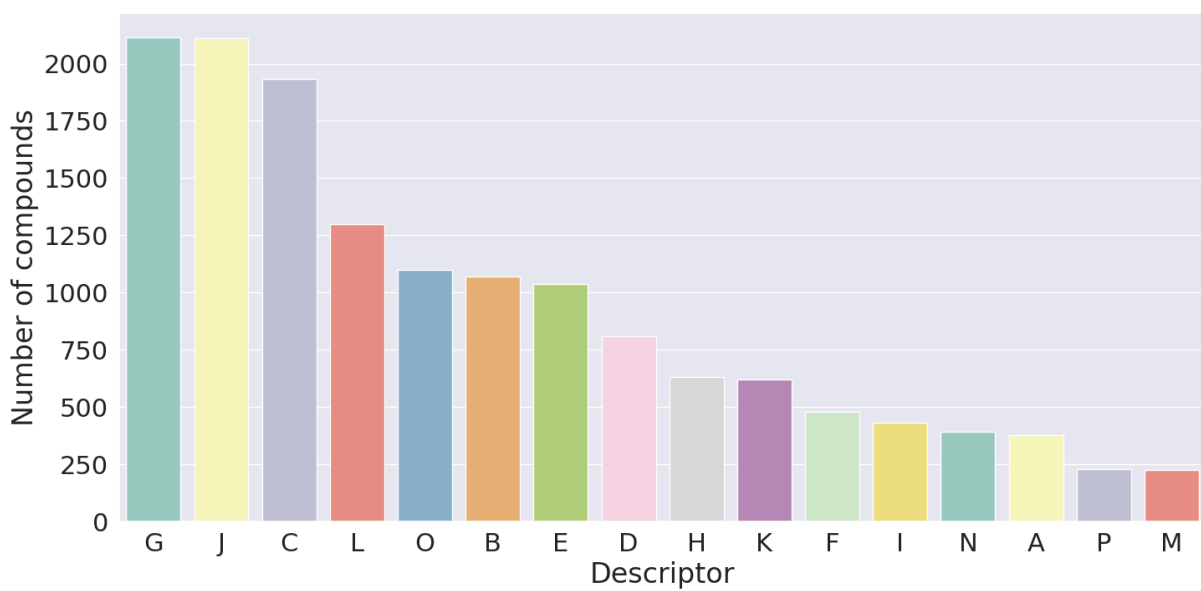

**Figure S10.** Number of compounds in the 16 different classes using the computer derived taxonomy. It should be noted that the classes are still imbalanced. See Figure S9 for the distribution of compounds in the 16 expert derived classes.

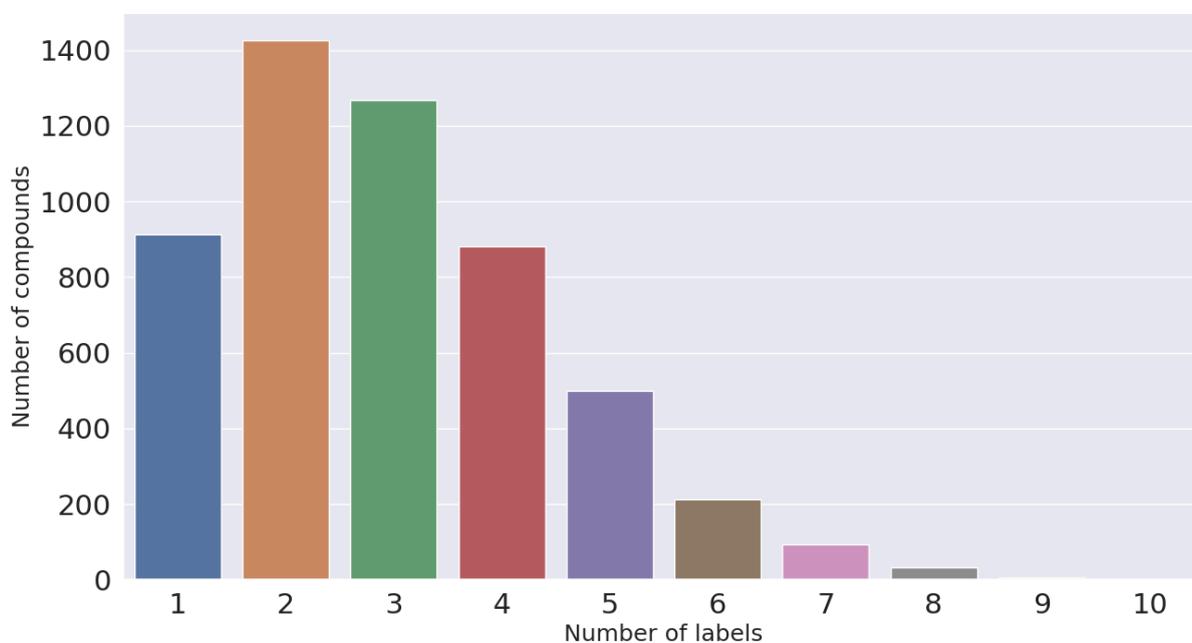

**Figure S11.** Overview of chemical compounds and the number of corresponding labels or descriptors provided for the dataset imposed with the expert derived taxonomy. The majority of compounds have between 1 and 4 labels similar to that of the computer taxonomy. Likewise, the upper bound of the possible number of labels has decreased from 15-17 in the dataset with all the descriptors to 8-10.

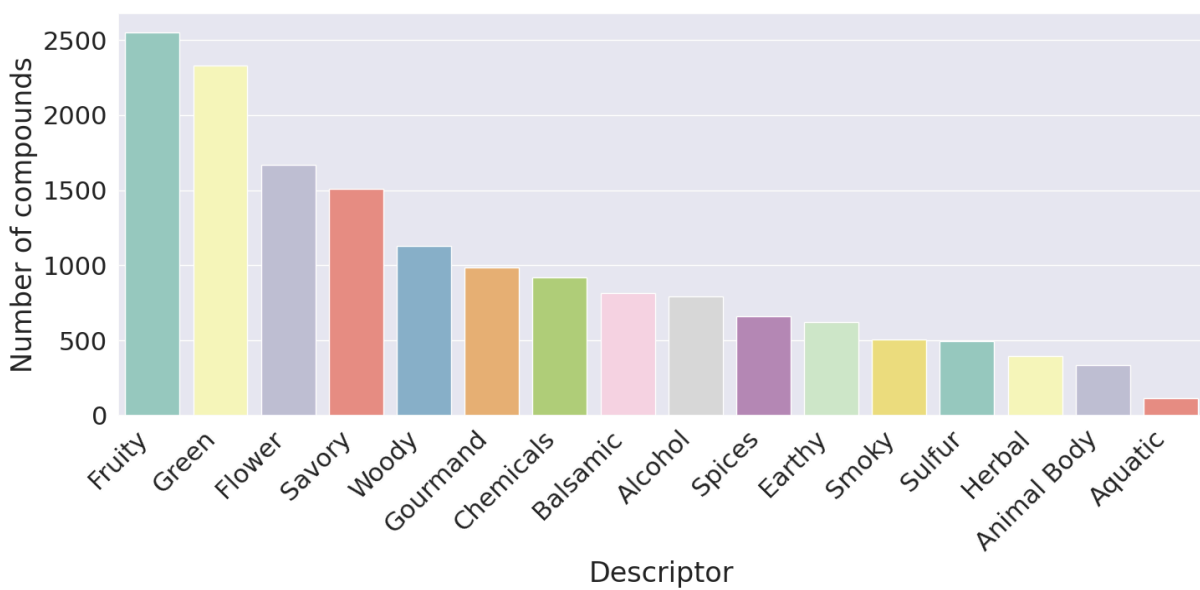

**Figure S12.** Number of compounds in each of the different classes, i.e., provided smell descriptors, for a total of 16 parent term descriptors of the expert taxonomy. It should be noted that the classes are still imbalanced.

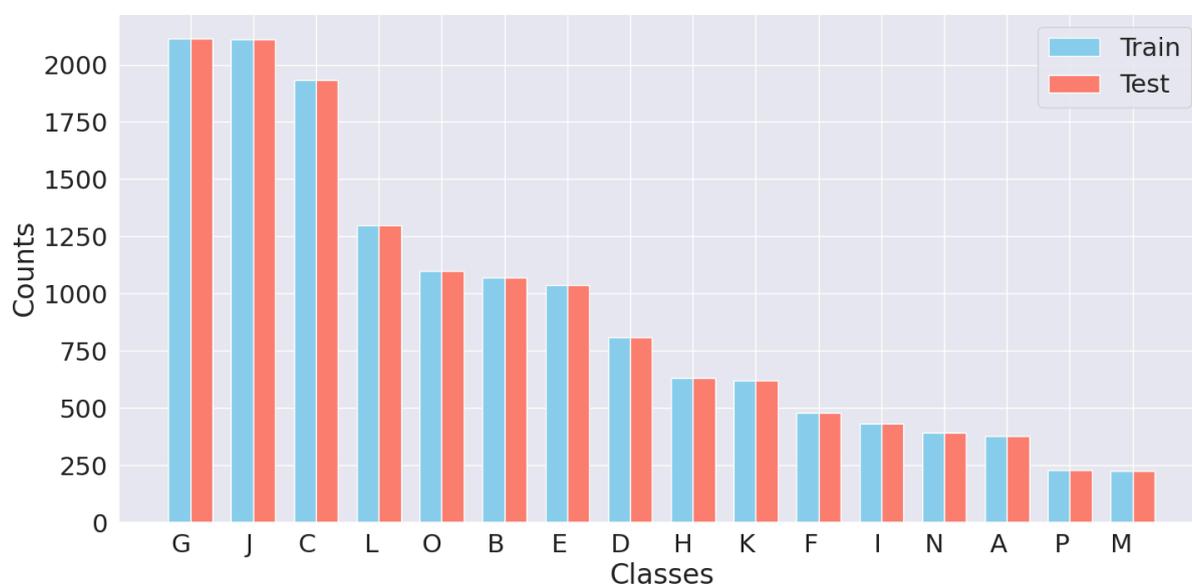

**Figure S13.** Number of compounds in each of the different classes using the computer taxonomy across the train and test splits.

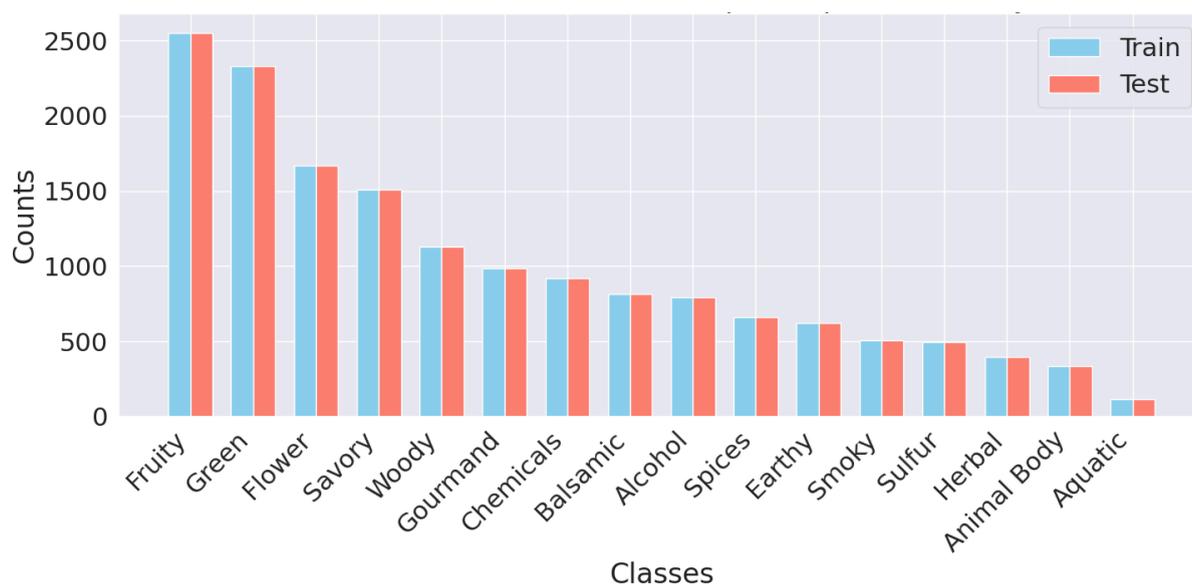

**Figure S14.** Number of compounds in each of the different classes using the expert taxonomy across both train and test splits.

**ERROR ANALYSIS AND FEATURE IMPORTANCE**

To investigate the most relevant features that the machine learning model uses for its predictions, the model's feature importance is looked at. However, since the feature importance of the model is based on how the model fits the training dataset, it's not as reliable. Permutation feature importance (PFI) proves a more reliable technique, where permutations are added to the features of the test data to rank features based on how much a given score metric drops upon permuting a given feature. The limitation of the PFI is that it looks at the drop of overall macro scores for evaluating feature importance and there is no class specificity. Therefore, SHAP Value Analysis are used in a second step (see further below).

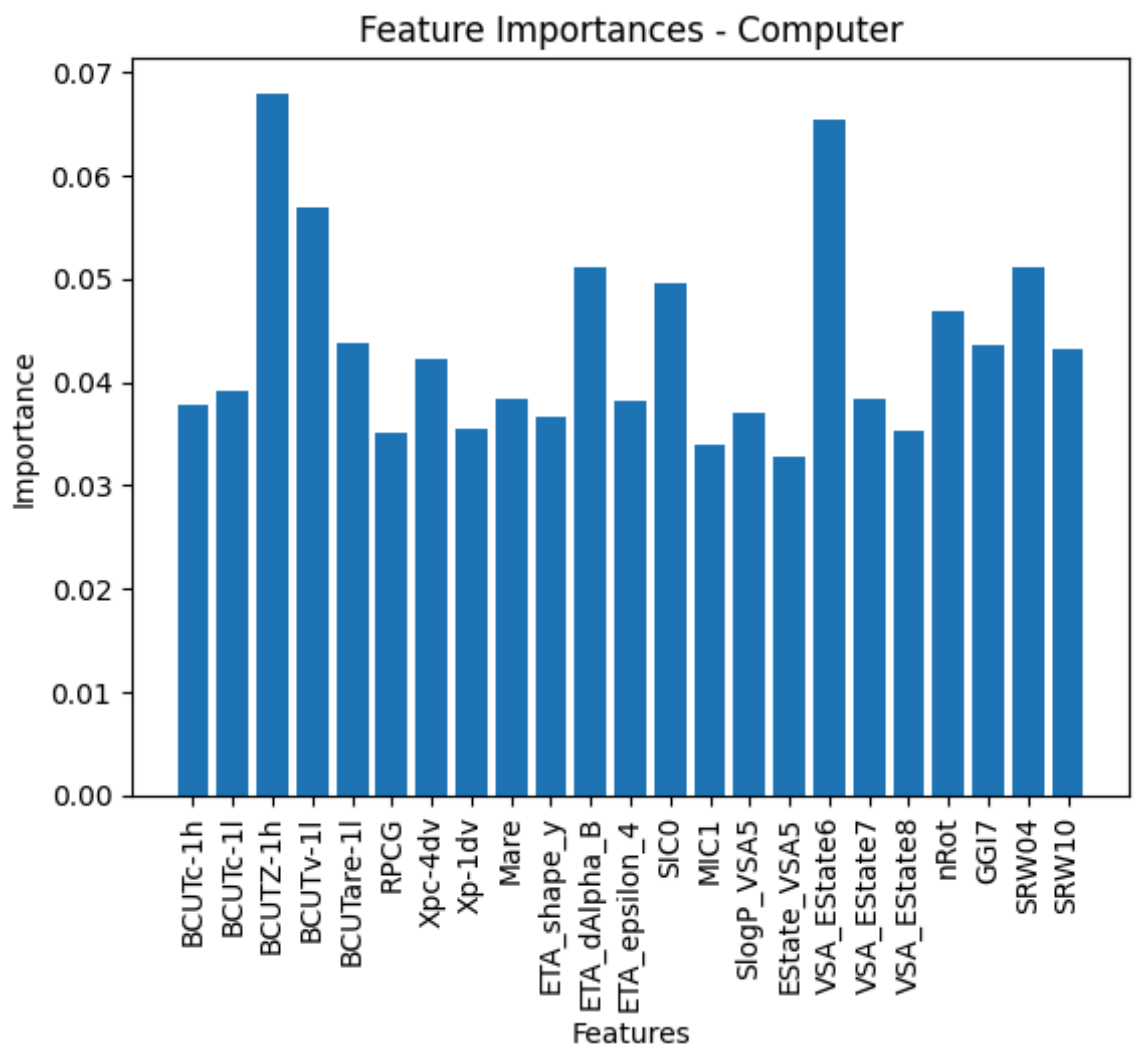

**Figure S15.** Feature importance of the XGBoost model for the Computer derived Taxonomy. The plot shows that the features BCUTZ-1h, BCUTv-1l and VSA-Estate6 are relevant for the model to complete the classification task. Note that this is based on the training data alone.

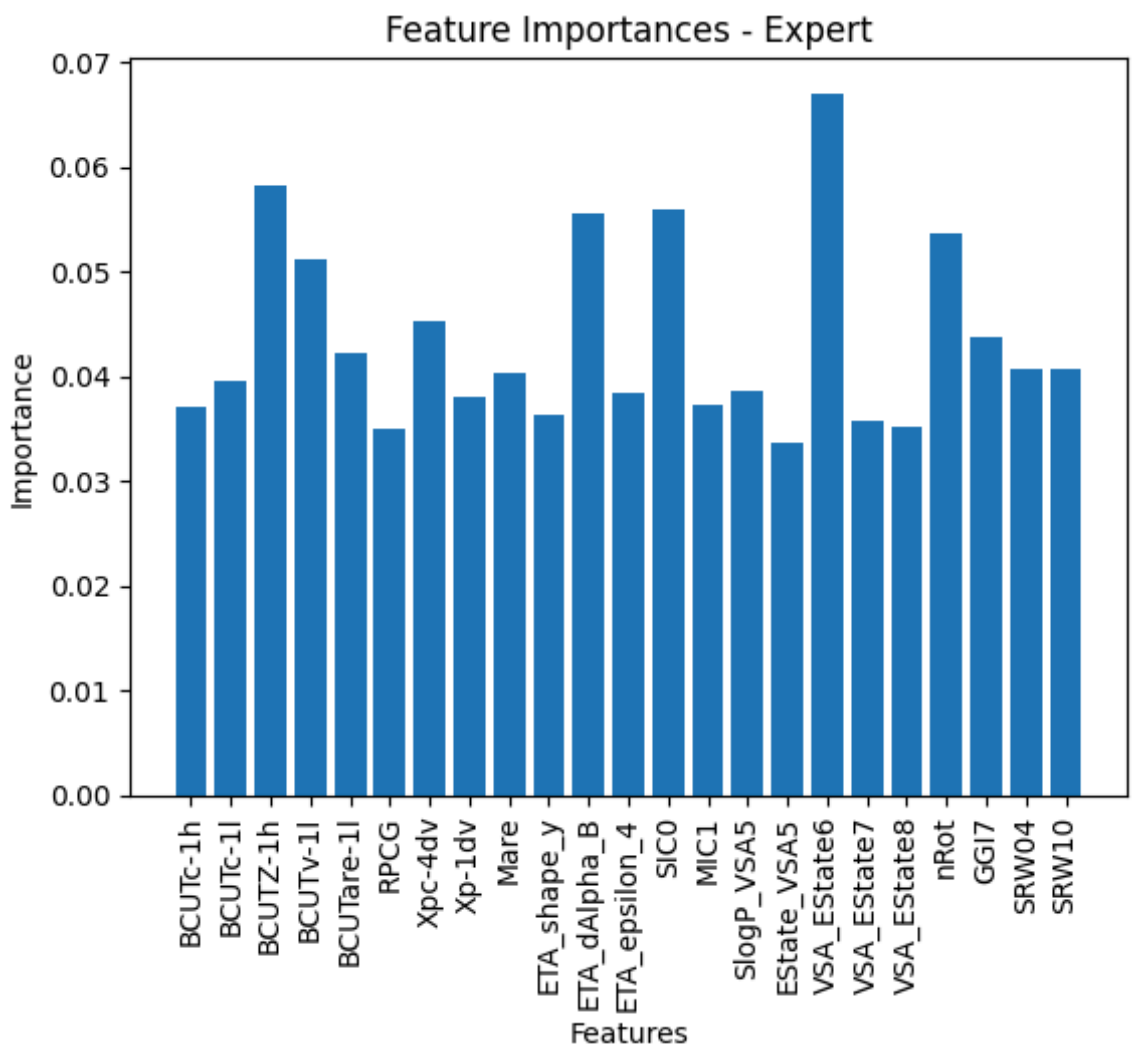

**Figure S16.** Feature importance of the XGBoost model for the Expert derived Taxonomy. The plot shows that the features BCUTZ-1h, SIC0 and VSA-Estate6 are relevant for the model to complete the classification task. Note that this is based on the training data alone.

**SHAP Value Analysis.** The SHapley Additive exPlanations (SHAP) analysis employs a game-theoretic approach to assign SHAP values, which represent the contribution or importance of each feature to a machine learning model's output [LUN]. The resulting values can be used to interpret the model predictions. We provide the SHAP summary plots for the different classes in both the computational and expert taxonomies. The features are listed on the y-axis, while the x-axis indicates the magnitude and direction of their contribution to the performance output of the XGBoost classifier. Each point in the plots represents an individual data instance, with color indicating the actual feature value (red for high, blue for low). Features are ranked based on importance, with those at the top contributing most significantly. The horizontal spread of points reflects the extent of a feature's influence across the dataset, with wider distributions imply greater overall impact on model predictions.

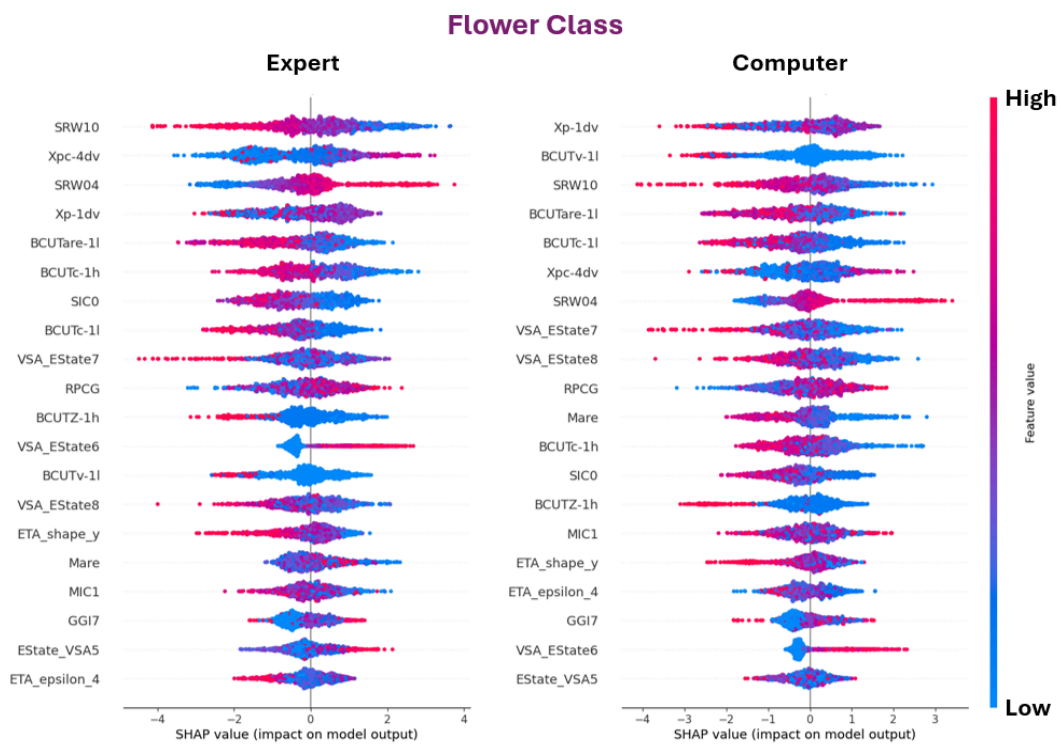

**Figure S17.** SHAP analysis of the "Flower" class for the XGBoost classifier in both taxonomies, see manuscript for details (right-hand side: computer derived; left-hand side expert derived).

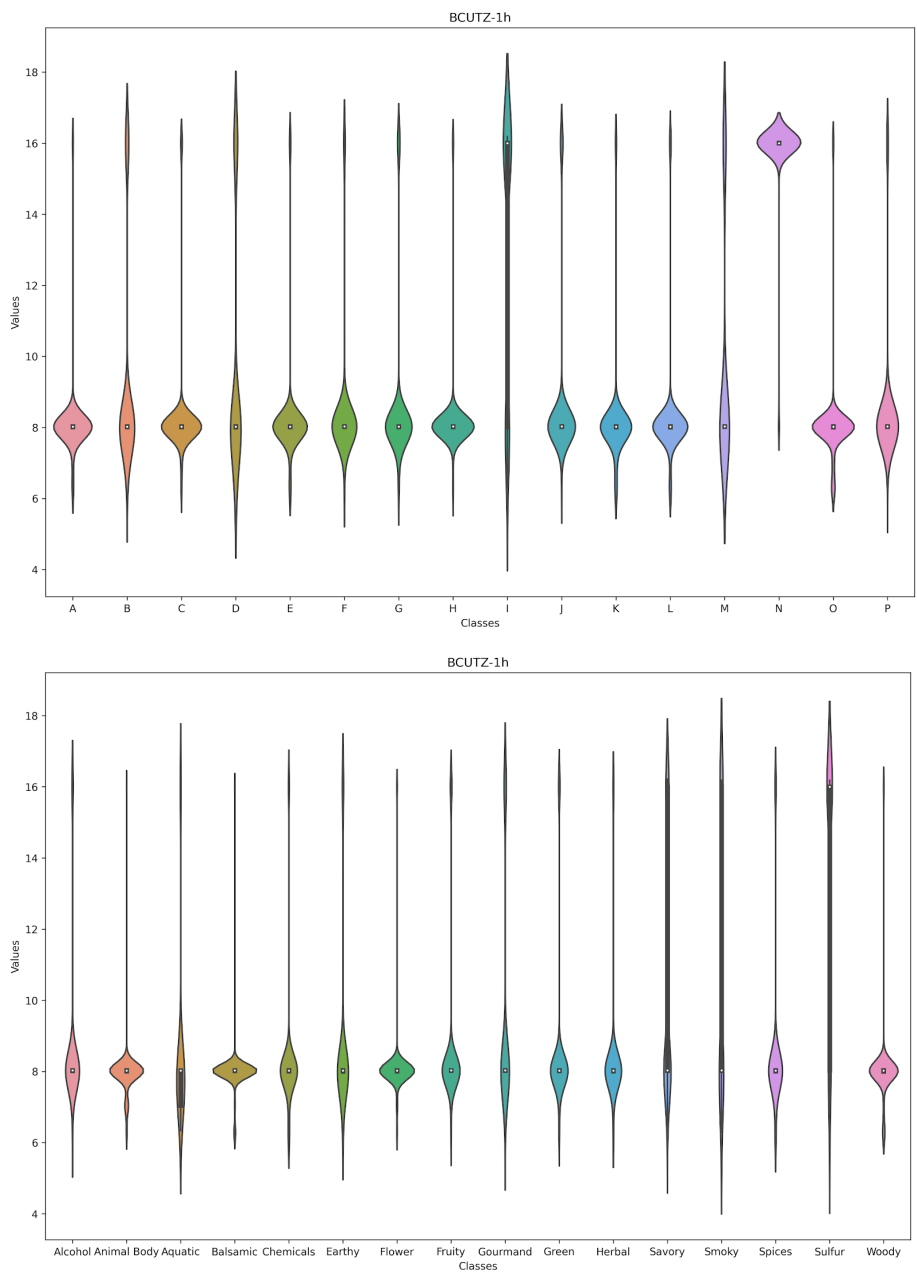

**Figure S18.** The feature 'BCUTZ-1h' across both taxonomies, computer and expert derived (upper and lower part, respectively). This is the most defining feature for the 'Sulfur' class across both taxonomies, corresponding to the atomic number of the atoms in the molecule samples, 16 in the case of sulfur atoms. For the Computer Taxonomy, we can see that a lot of data points cluster around 16 for the 'N' class which conceptually corresponds to the Sulfur class, as well as for the 'I' class, which corresponds to the "Savory" class of the expert taxonomy.

The remaining pages consist of classwise SHAP plots for all classes across both taxonomies which individually show feature contributions for the model's decision making.

**Figure S19:** SHAP analysis of the XGBoost classifier on the computer taxonomy for all 16 classes (pages 24-27). See also manuscript for the available code on the GitHub repository.

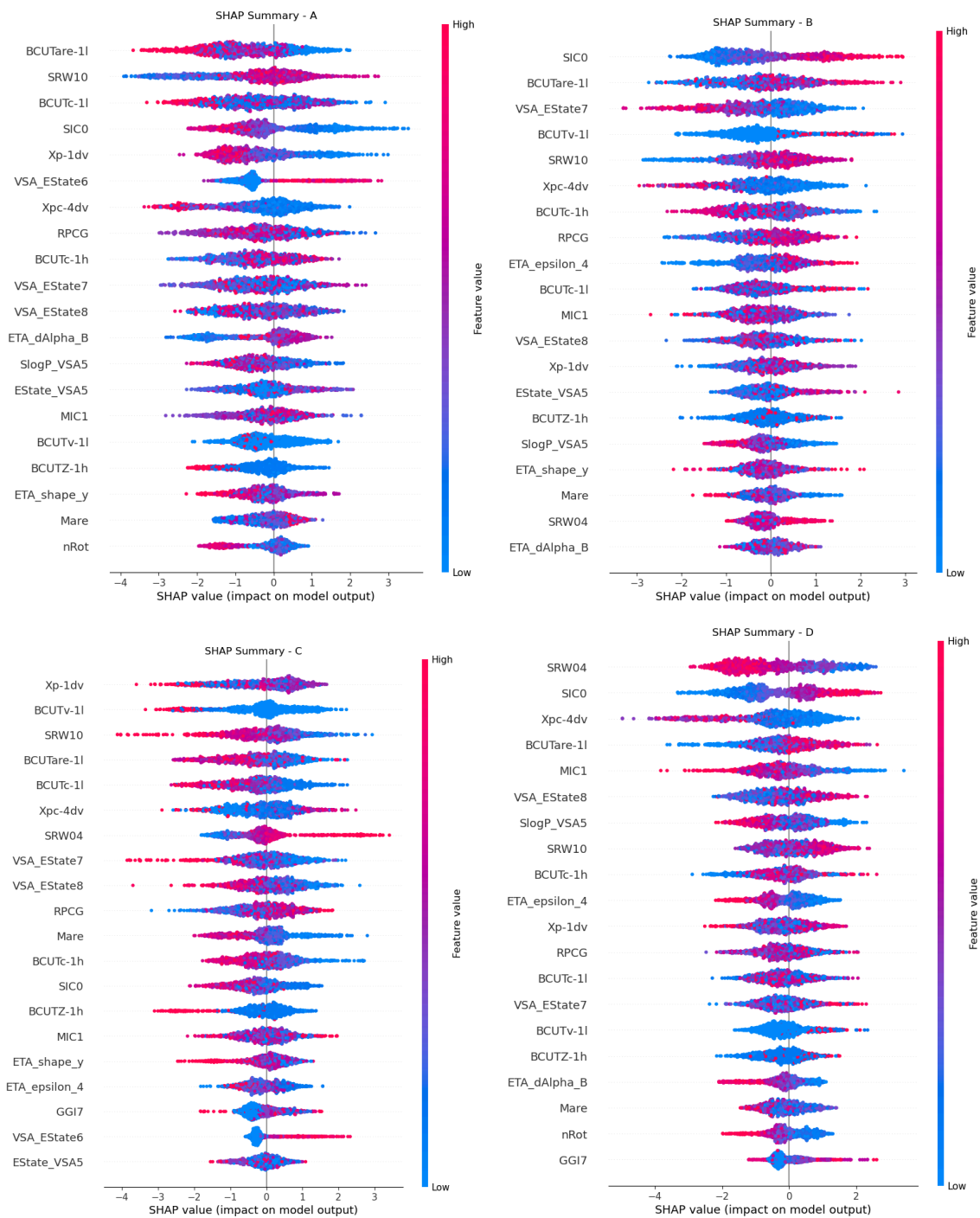

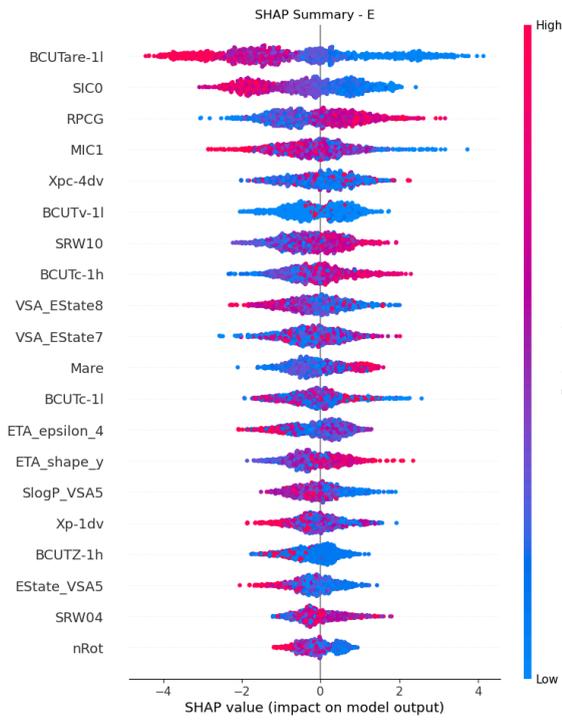
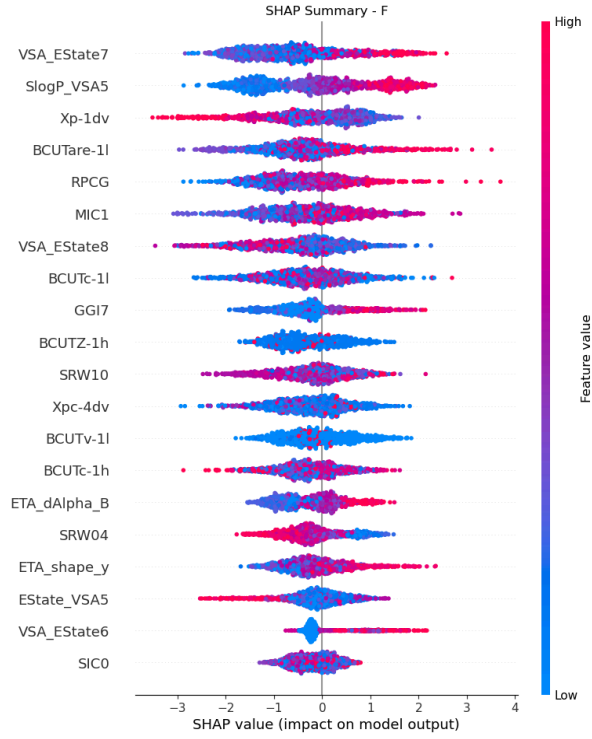
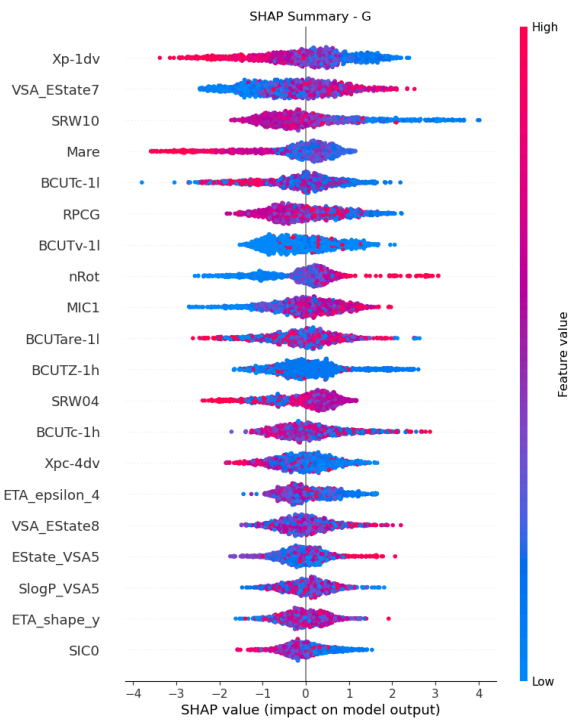
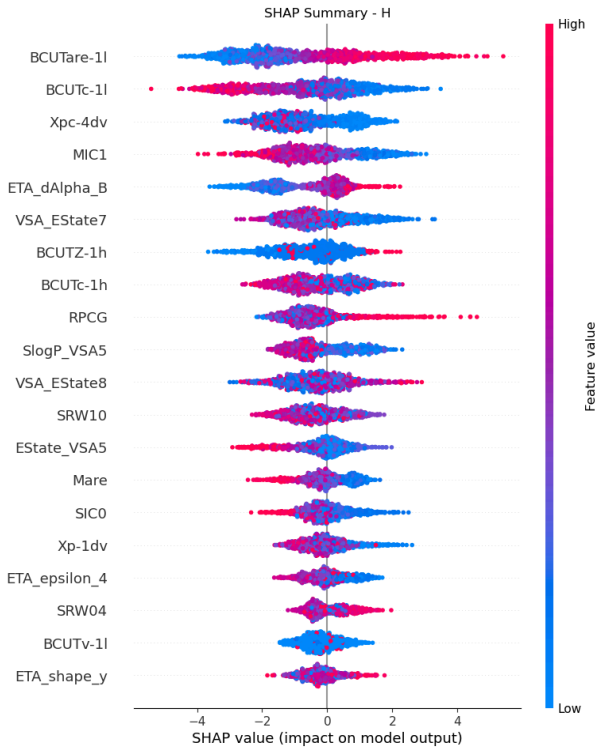

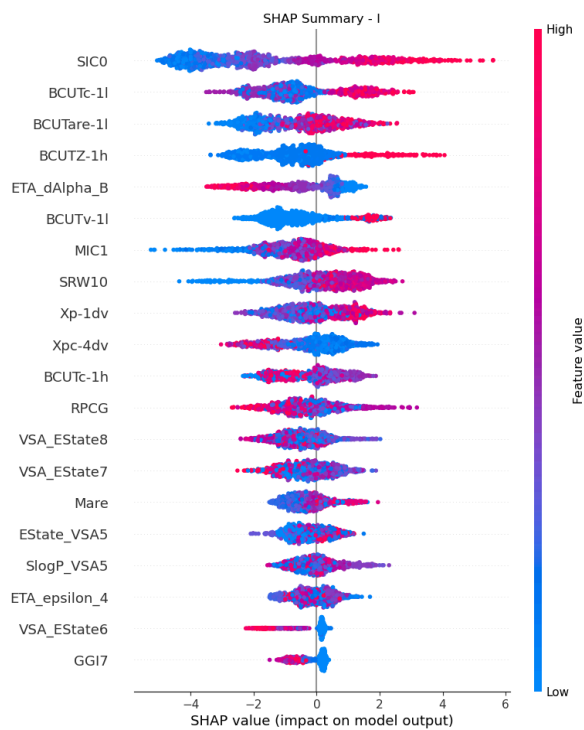
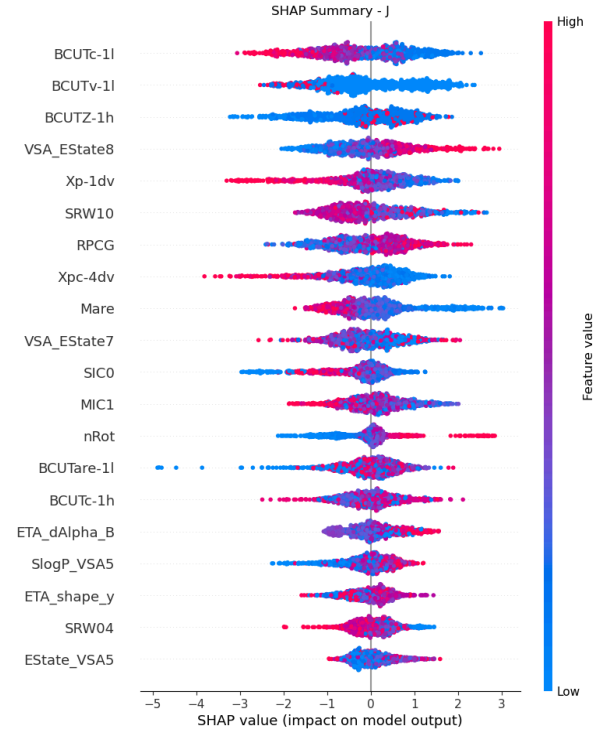
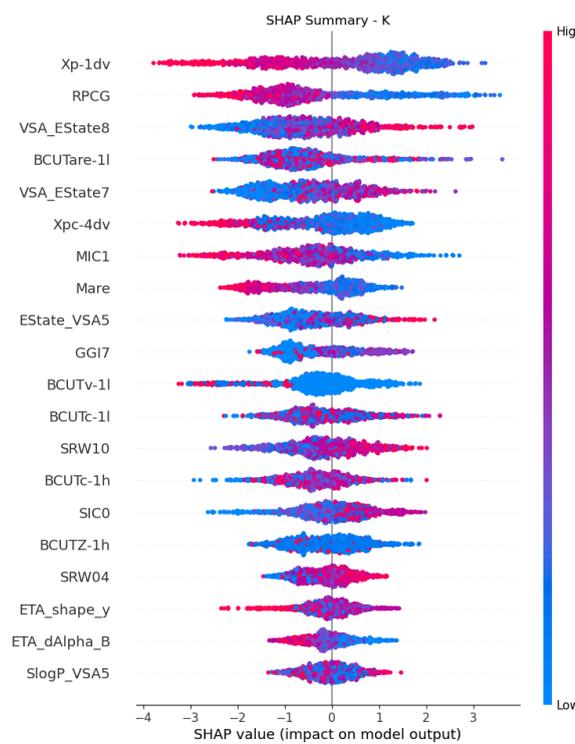
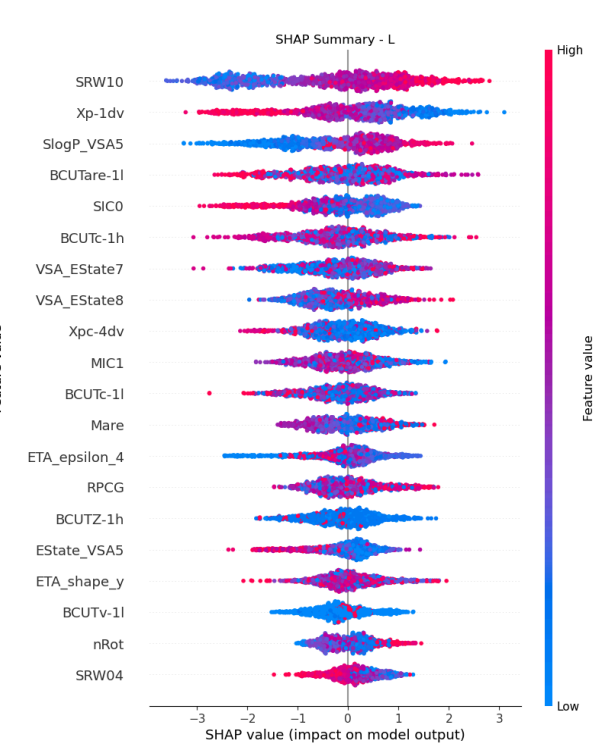

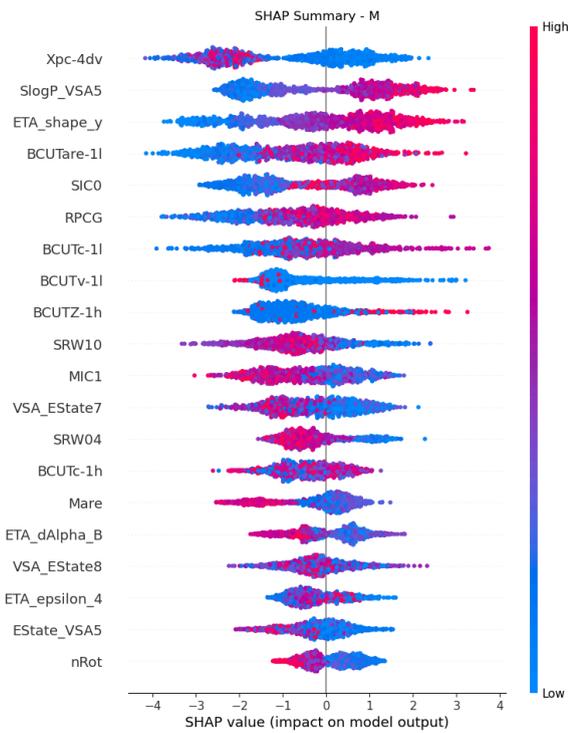
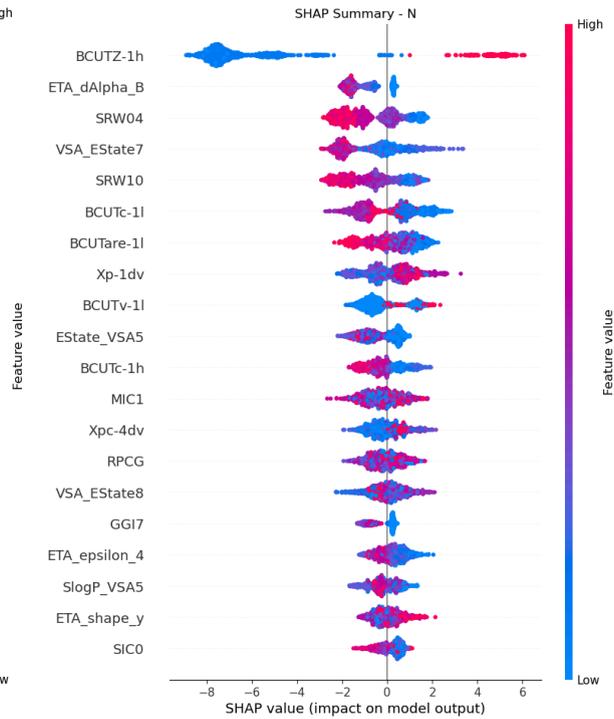
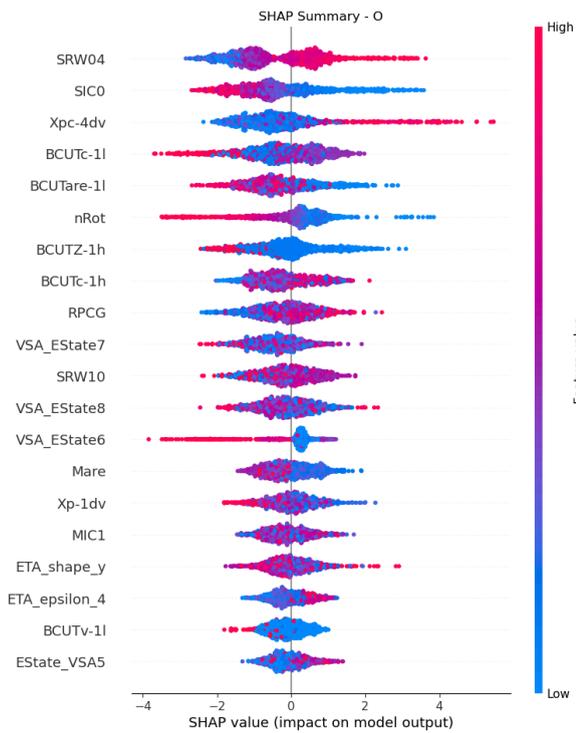
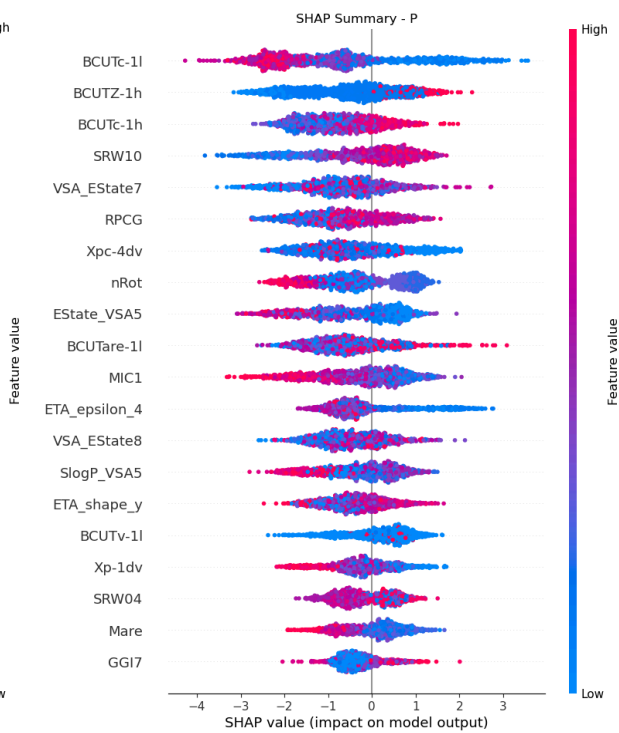

**Figure S20.** SHAP analysis of the XGBoost classifier on the expert taxonomy for all 16 classes (pages 28-31). See also manuscript for the available code on the GitHub repository.

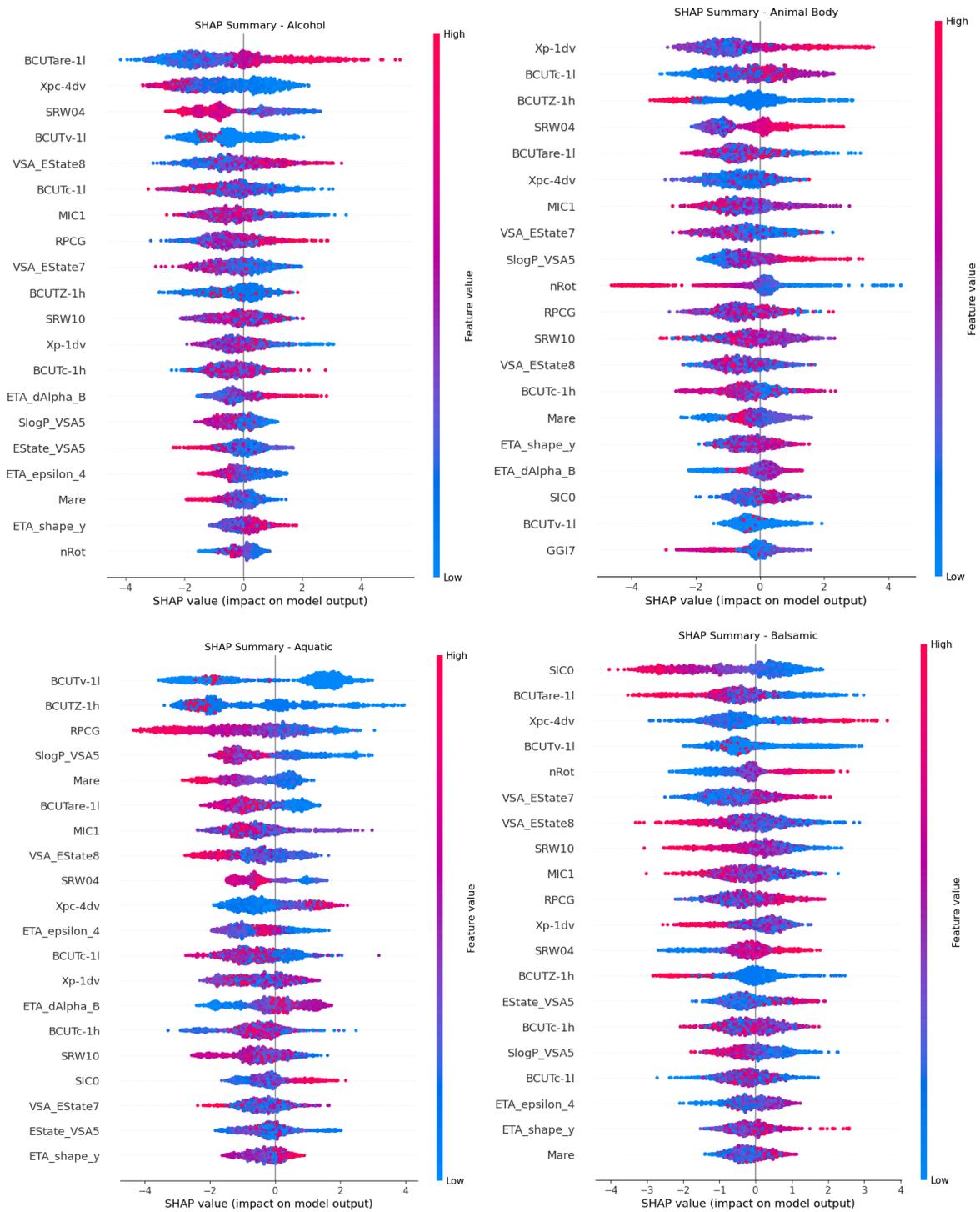

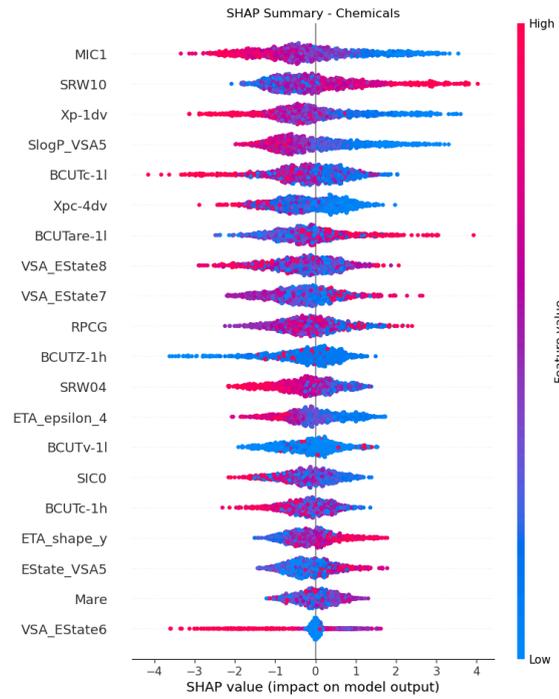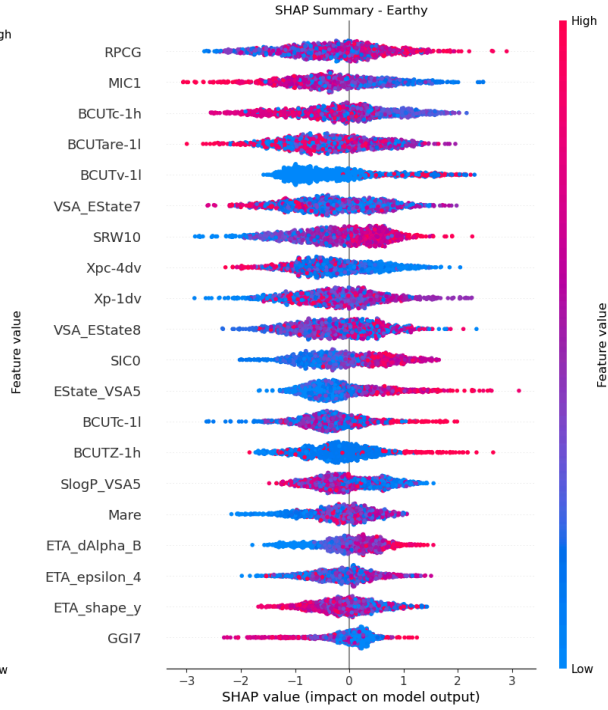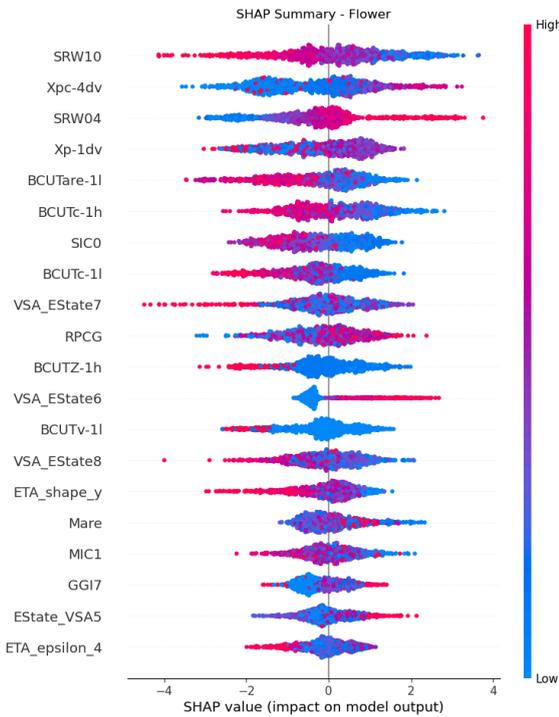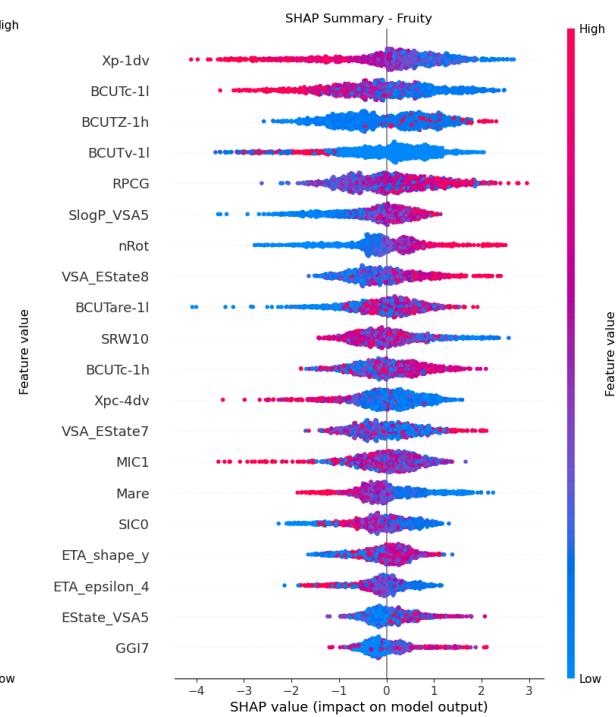

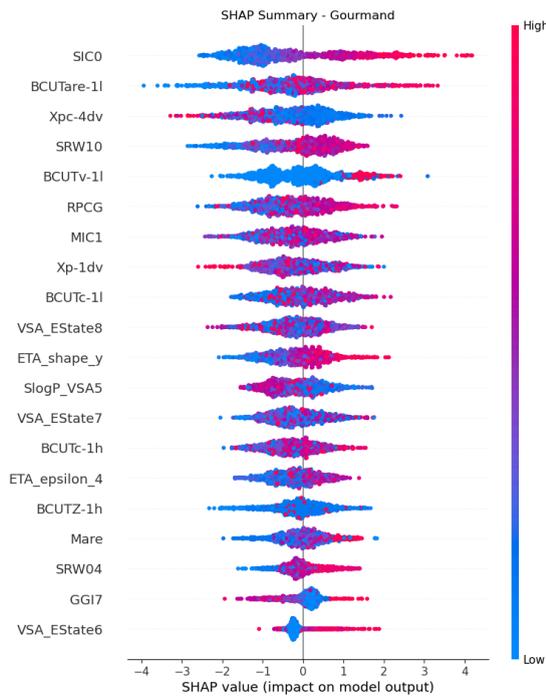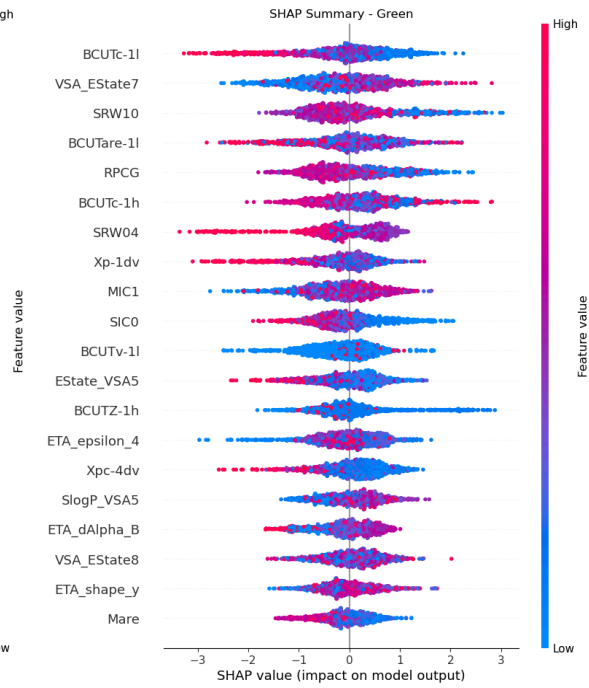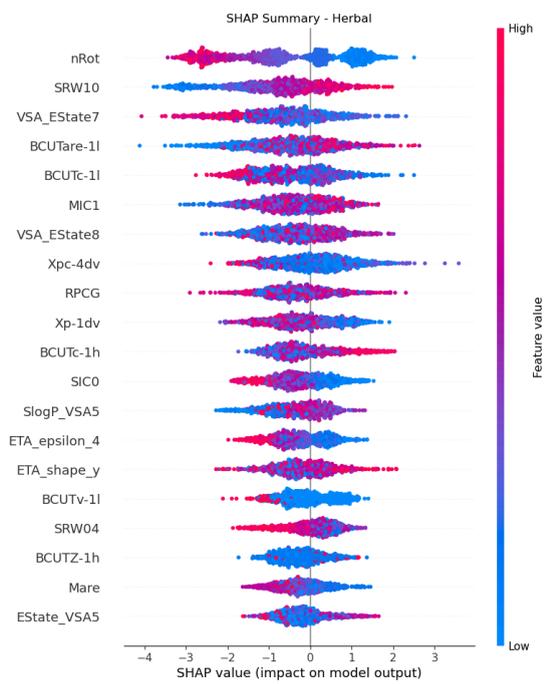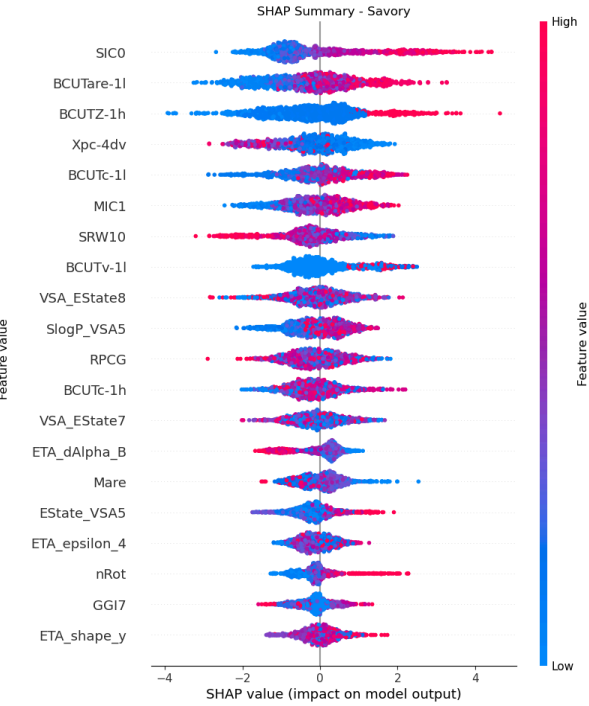

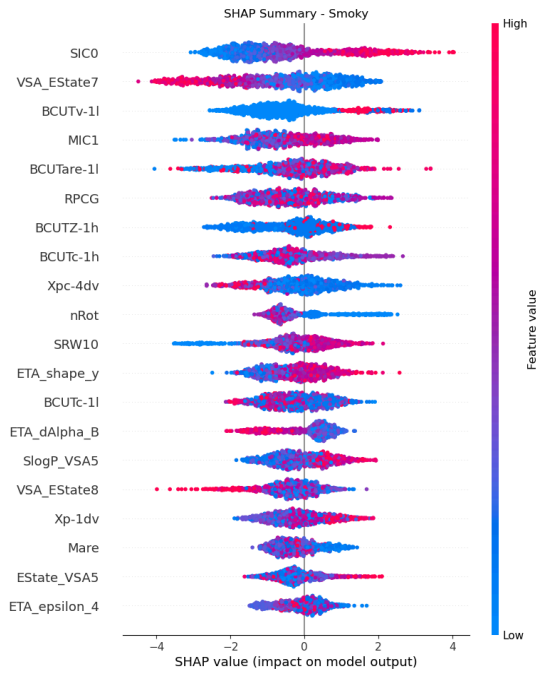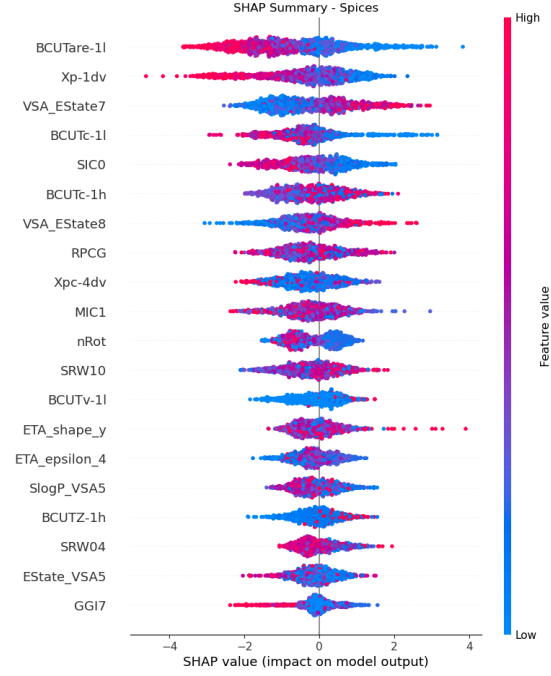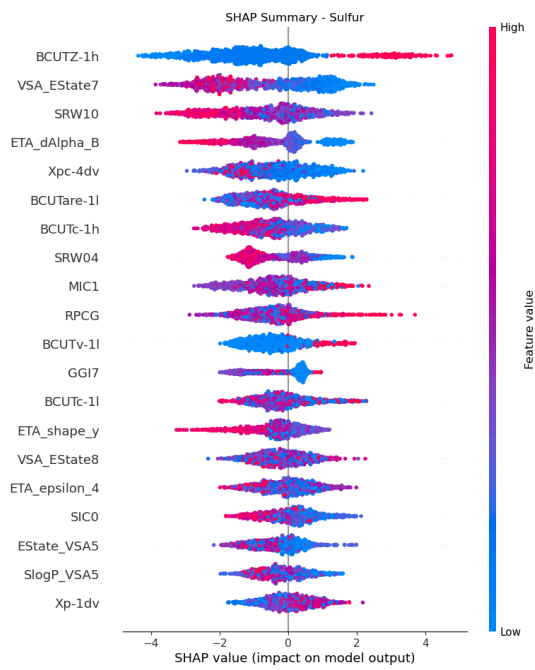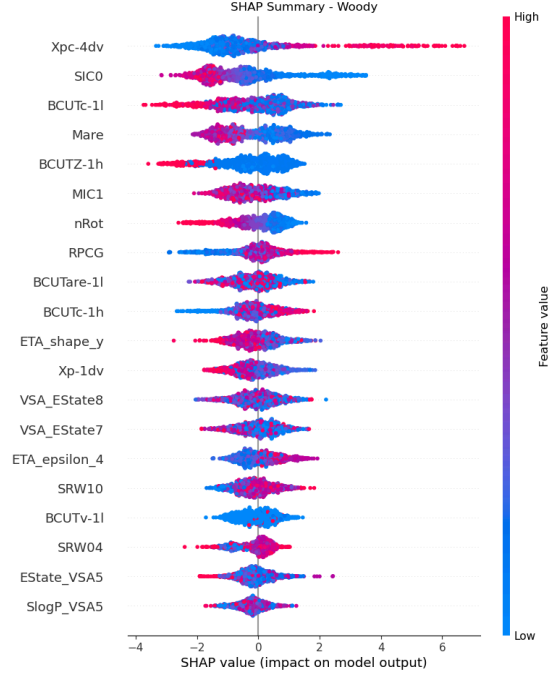